\documentclass[11pt]{article}
\usepackage{fullpage}
\usepackage{rotating}

\usepackage{amssymb,amsmath}
\usepackage{graphicx, epsfig}

\def\idrm#1{\ensuremath{\mathrm{#1}}}

\newcommand{\no}[1]{}

\newtheorem{theorem}{Theorem}
\newtheorem{lemma}{Lemma}

{\unskip\nobreak\hskip 1em plus 1fil\nobreak$\Box$
\parfillskip=0pt%
\endtrivlist}

\newcommand{\cT}{{\cal T}}
\newcommand{\cP}{{ K}}

\newcommand{\oS}{\overline{S}}

\newcommand{\oL}{\overline{L}}
\newcommand{\oB}{\overline{B}}
\newcommand{\oP}{\overline{{ K}}}
\newcommand{\oi}{\overline{i}}

\newcommand{\todo}[1]{ }

\newcommand{\ra}{\idrm{rank}}
\newcommand{\sel}{\idrm{select}}
\newcommand{\acc}{\idrm{access}}

\newcommand{\eps}{\varepsilon}

\begin{document}

\title{\Large Optimal Dynamic Sequence Representations
\thanks{Partially funded by Fondecyt grant 1-110066, Chile.
An early partial version of this paper appeared in {\em Proc. SODA'13}.}}

\author{
Gonzalo Navarro\thanks{Department of Computer Science, University of Chile.
{\tt  gnavarro@dcc.uchile.cl}.}
\and
Yakov Nekrich\thanks{Department of Electrical Engineering and Computer Science,
University of Kansas.
{\tt yakov.nekrich@googlemail.com}.}
}
\date{}

\maketitle

\begin{abstract} 
We describe a data structure that supports access, rank and select queries,
as well as symbol insertions and deletions, on a string $S[1,n]$ over alphabet 
$[1..\sigma]$ in time $O(\lg n/\lg\lg n)$, which is optimal even on binary
sequences and in the amortized sense. Our time is worst-case for the queries 
and amortized for the updates. This complexity is better than the best previous
ones by a $\Theta(1+\lg\sigma/\lg\lg n)$ factor.
We also design a variant where times are worst-case, yet rank and updates
take $O(\lg n)$ time.
Our structure uses $nH_0(S)+o(n\lg\sigma) + O(\sigma\lg n)$
bits, where $H_0(S)$ is the zero-order entropy of $S$.
\no{Finally, we pursue various extensions, like handling general alphabets
(such as reals and strings) and supporting a more general algebra of string
operations including concatenations, splits, and block edits.}
Finally, we pursue various extensions and applications of the
result.
\end{abstract}

\section{Introduction}
\label{sec:intro}

String representations supporting $\ra$ and $\sel$ queries are fundamental
in many data structures, including 
full-text indexes \cite{GGV03,FMMN07,GMR06}, 
permutations \cite{GMR06,BGNN10}, 
inverted indexes \cite{BFLN08,BGNN10}, 
graphs \cite{CN10},
document retrieval indexes \cite{VM07}, 
labeled trees \cite{GMR06,BHMR07}, 
XML indexes \cite{GHSV07,FLMM09},
binary relations \cite{BHMR07}, 
and many more. 
The problem is to encode a string $S[1,n]$ over alphabet $\Sigma=[1..\sigma]$ 
so as to support the following queries:
\begin{eqnarray*}
\ra_a(S,i) &=& \textrm{\small number of occurrences of $a \in \Sigma$ } 
            \textrm{\small in $S[1,i]$, for $1 \le i \le n$.} \\
\sel_a(S,i) &=& \textrm{\small position in $S$ of the $i$-th occurrence } 
             \textrm{\small of $a \in \Sigma$, for $1 \le i \le \ra_a(S,n)$.} \\
\acc(S,i) &=& S[i].
\end{eqnarray*}

There exist various {\em static} representations of $S$ (i.e., $S$ cannot change) that support these 
operations \cite{GGV03,GMR06,FMMN07,BGNN10,BN12}. The most recent one
\cite{BN12} shows a lower bound of $\Omega(\lg\frac{\lg\sigma}{\lg w})$ time
for operation $\ra$ on a RAM machine with $w$-bit words, using any
$O(n \lg^{O(1)} n)$ space. It also provides a matching upper bound that in addition 
achieves almost constant time for $\sel$ and $\acc$, using compressed space. 
Thus the problem is essentially closed.

However, various applications need dynamism, that is, the ability to insert and 
delete symbols in $S$. A lower bound for this case, in order to support
operations $\ra$, insertions and deletions, even for bitmaps ($\sigma=2$) and
in the amortized sense, is $\Omega(\lg n /\lg\lg n)$ \cite{FS89}. On the other
hand the best known upper bound \cite{HM10,NS10} is 
$O((1+\lg\sigma/\lg n)\lg n /\lg\lg n)$, that is, a factor 
$\Theta(\lg\sigma/\lg\lg n)$ away from the lower bound for
alphabets larger than polylogarithmic. Their space is $nH_0(S)+o(n\lg\sigma)$,
where $H_0(S) = \sum_{a \in [1..\sigma]} (n_a/n) \lg(n/n_a) \le \lg\sigma$
is the zero-order entropy of $S$, where $n_a$ is the number of occurrences of 
$a$ in $S$. 

In this paper we close this gap by providing an {\em optimal-time} dynamic
representation of sequences. Our representation takes $O(\lg n /\lg\lg n)$
time for all the operations, worst-case for the three queries and amortized
for insertions and deletions. We present a second variant achieving worst-case
bounds for all the operations, $O(\lg n /\lg\lg n)$ for $\sel$ and $\acc$
and $O(\lg n)$ for $\ra$, insertions and deletions. The space is also
$nH_0(S)+o(n\lg\sigma)$. This $O(\lg n)$ is still faster than previous
work for $\lg\sigma = \Omega((\lg\lg n)^2)$. This gets much closer to
closing this problem under the dynamic scenario as well.

We then 
\no{consider some extensions of our results. First, we }
show how to 
handle general alphabets, such as $\Sigma= \mathbb{R}$, or 
$\Sigma=\Gamma^*$ for a symbol alphabet $\Gamma$, in optimal time. 
For example, in the comparison model for $\Sigma=\mathbb{R}$,
the time is $O(\lg\sigma+\lg n / \lg\lg n)$, where $\sigma$ is the number of 
distinct symbols that appear in $S$; in the case $\Sigma=\Gamma^*$ for 
general $\Gamma$, the time is $O(|a|+\lg\gamma + \lg n / \lg\lg n)$, where
$|a|$ is the length of the involved symbol (a string) and $\gamma$ the number of distinct 
symbols of $\Gamma$ that appear in the elements of $S$.
Previous dynamic solutions have assumed that the alphabet
$[1..\sigma]$ was static.

\no{ -> needs quadratic blocks, at best, but how to handle the split-find
and partial sums?
As a second extension, we enhance the set of update operations to include
sequence splitting, concatenating, and general block updates. This makes up
a much stronger algebra on sequences. The time for these operations is
$O(\sigma\lg n)$, whereas the only previous solution \cite{NS10} achieved
$O(\sigma\lg^{1+\eps} n)$ for any constant $\epsilon>0$.
}

At the end we describe several applications where our result offers improved
time/space tradeoffs. These include compressed indexes for dynamic text 
collections, construction of the Burrows-Wheeler transform \cite{BW94} and 
static compressed text indexes within compressed space, as well as compressed
representations of dynamic binary relations, directed graphs, and inverted
indexes.
 
We start with an overview of the state of the art, putting our solution in 
context, in Section~\ref{sec:related}. We review the wavelet tree data 
structure \cite{GGV03}, which is at the core of our solution (and most
previous ones) in Section~\ref{sec:wavel}. In Section~\ref{sec:basic} we 
describe the core of
our amortized solution, deferring to Section~\ref{sec:rank} the management of 
deletions and its relation with a split-find data structure needed for rank
and inserts. Section~\ref{sec:space} deals with the changes in $\lg n$ and
how we obtain times independent of $\sigma$, and concludes
with Theorem~\ref{thm:optimal}, our result on uncompressed sequences.
Then Section~\ref{sec:rrr} shows how to improve the data encoding to obtain
compressed space in Theorem~\ref{thm:compr}, and Section~\ref{sec:worst} 
how to obtain worst-case times, Theorem~\ref{thm:worstcase}.
\no{Finally, Section~\ref{sec:ext} describes some extensions of our results and
Section~\ref{sec:app} describes various applications. }
Finally, Section~\ref{sec:app} describes some extensions and several applications of our results.
We conclude in
Section~\ref{sec:concl}. An important technical part of the paper, describing
the structure of blocks that handle subsequences of polylogarithmic size, is
deferred to Section~\ref{app:A} to avoid distractions.

\section{Related Work}
\label{sec:related}

With one exception \cite{GHSV07}, all the previous work on dynamic sequences
build on the {\em wavelet tree} structure \cite{GGV03}.
The wavelet tree decomposes $S$ hierarchically. In a first level, it separates
larger from smaller symbols, by marking in a bitmap which symbols of $S$ were
larger and which were smaller. The two subsequences of $S$ are recursively
separated. The $\lg\sigma$ levels of bitmaps describe $S$, and $\acc$, $\ra$
and $\sel$ operations on $S$ are carried out via $\lg\sigma$ $\ra$ and
$\sel$ operations on the bitmaps (see Section~\ref{sec:wavel} for more details).

In the static case, $\ra$ and $\sel$ operations on bitmaps take constant time,
and therefore $\acc$, $\ra$ and $\sel$ on $S$ takes $O(\lg\sigma)$ time
\cite{GGV03}. This can be reduced to $O(1+\lg\sigma/\lg\lg n)$ by using
multiary wavelet trees \cite{FMMN07}. These separate the symbols into
$\rho=o(\lg n)$ ranges, and instead of a bitmap store a sequence over an 
alphabet of size $\rho$. 

Insertions and deletions in $S$ can also be carried out by inserting and 
deleting bits from $\lg\sigma$ bitmaps. However, the operations on dynamic 
bitmaps are bound to be slower. Fredman and Saks \cite{FS89} show that 
$\Omega(\lg n /\lg\lg n)$ time is necessary, even in the amortized sense, to 
support $\ra$, insert and delete operations on a bitmap. By using dynamic 
bitmap solutions \cite{HSS03,CHL04,BB04,CHLS07,HSS11} on the wavelet tree
levels, one immediately obtains a dynamic sequence representation, 
where the space and the time of the dynamic bitmaps
solution is multiplied by $\lg\sigma$ (the sum of the zero-order entropies of
the bitmaps also adds up to $nH_0(S)$ \cite{GGV03}). With this combination one
can obtain times as good as $O(\lg\sigma\lg n /\lg\lg n)$ (and $n\lg\sigma +
o(n\lg\sigma)$ bits) \cite{CHLS07} and spaces as good as $O(nH_0(S))$ 
(and $O(\lg\sigma\lg n)$ time) \cite{BB04}.

M\"akinen and Navarro \cite{MN06,MN08} made the above combination explicit
and obtained compressed bitmaps with logarithmic-time operations, yielding
$O(\lg\sigma \lg n)$ time for all the sequence operations and the best
compressed space until then, $nH_0(S)+o(n\lg\sigma)$ bits. 
They also obtained $O((1+\lg\sigma/\lg\lg n)\lg n)$ query time, but with
an update time of $O(\lg\sigma \lg^{1+\eps} n)$, for any constant $0<\eps<1$.
This was achieved by replacing binary with multiary wavelet trees, and
obtaining $O(\lg n)$ query time for the operations on sequences over a small
alphabet of size $o(\lg n)$.

Lee and Park \cite{LP07,LP09} pursued this path further, obtaining the
$O(1+\lg\sigma/\lg\lg n)\lg n)$ time for queries and update
operations, yet the space was not compressed, $n\lg\sigma+o(n\lg\sigma)$,
and update times were amortized.
Shortly after, Gonz\'alez and Navarro \cite{GN08,GN09} obtained the best
of both worlds, making all the times worst-case and compressing the space
again to $nH_0(S)+o(n\lg\sigma)$ bits. Both solutions managed to solve all
query and update operations in $O(\lg n)$ time on sequences over small
alphabets of size $o(\lg n)$.

Finally, almost simultaneously, He and Munro \cite{HM10} and Navarro and 
Sadakane \cite{NS10} obtained the currently best result, 
$O((1+\lg\sigma/\lg\lg n)\lg n /\lg\lg n)$ time, still within the same 
compressed space. They did so by improving the times of
the dynamic sequences on small alphabets to $O(\lg n / \lg\lg n)$, which is
optimal even on bitmaps and in the amortized sense \cite{FS89}. 

As mentioned, the solution by Gupta et al.~\cite{GHSV07} deviates from this
path and is a general framework for using any static data structure
and periodically rebuilding it. By using it over a given representation
\cite{GMR06}, it achieves $O(\lg\lg n)$ query time and $O(n^\eps)$ amortized
update time. It would probably achieve compressed space if combined with more
recent static data structures \cite{BGNN10}. This shows that query times can
be significantly smaller if one allows for much higher update times. In this
paper, however, we focus in achieving similar times for all the
operations.

\begin{sidewaystable}[p]
\caption{History of results on managing dynamic sequences $S[1,n]$ over alphabet
$[1..\sigma]$, assuming $\sigma=o(n/\lg n)$ to simplify. 
Some results \cite{HSS03,BB04,CHL04} were presented only for binary 
sequences and the result we give is obtained 
by using them in combination with wavelet trees.
Column W/A tells whether the update times are (W)orst-case or (A)mortized.}
\label{tab:res}
\begin{center}
\footnotesize
\begin{tabular}{l|c|c|c|c}
Source & Space (bits) & Query time & Update time & W/A \\
\hline
\cite{FS89} &    & \multicolumn{2}{c|}{$\Omega(\lg n /\lg\lg n)$ for $\ra$ and indels} & A \\
\hline
\cite{HSS03,HSS11} & $n\lg\sigma+O(n\lg\sigma(\lg\lg n)^2/\lg n)$ & $O(\lg\sigma\lg n/\lg\lg n)$ & $O(\lg\sigma(\lg n /\lg\lg n)^2)$ & A \\
\cite{CHL04} & $O(n\lg\sigma)$ & $O(\lg\sigma\lg n)$ & $O(\lg\sigma\lg n)$ & W \\
\cite{BB04} & $O(nH_0(S)+\lg n)$ & $O(\lg\sigma \lg n)$ & $O(\lg\sigma \lg n)$ & W \\
\cite{MN06,MN08} & $nH_0(S)+O(n\lg\sigma/\sqrt{\lg n})$ & $O(\lg\sigma\lg n)$ & $O(\lg \sigma \lg n)$ & W \\
            & $nH_0(S)+O(n\lg\sigma/\lg^{1/2-\eps} n)$ & $O((1+\frac{1}{\eps}\lg\sigma/\lg\lg n)\lg n)$ & $O(\frac{1}{\eps}\lg\sigma\lg^{1+\eps} n)$ & W \\
\cite{CHLS07} & $O(n\lg\sigma)$ & $O(\lg\sigma \lg n/\lg\lg n)$ & $O(\lg\sigma \lg n/\lg\lg n)$ & W \\
\cite{GHSV07} & $n\lg\sigma+O(n\lg\sigma/\lg\lg\sigma)$ & $O(\frac{1}{\eps}\lg\lg n+\lg\lg\sigma)$ & $O(\frac{1}{\eps}n^\eps)$ & A \\
\cite{LP07,LP09} & $n\lg\sigma+O(n\lg\sigma/\sqrt{\lg n})+O(n)$ & $O((1+\lg\sigma/\lg\lg n)\lg n)$ & $O((1+\lg\sigma/\lg\lg n)\lg n)$ & A \\
\cite{GN08,GN09} & $nH_0(S)+O(n\lg\sigma/\sqrt{\lg n})$ & $O((1+\lg\sigma/\lg\lg n)\lg n)$ & $O((1+\lg\sigma/\lg\lg n)\lg n)$ & W \\
\cite{HM10} & $nH_0(S)+O(n\lg\sigma/\sqrt{\lg n})$ & $O((1+\lg\sigma/\lg\lg n)\lg n/\lg\lg n)$ & $O((1+\lg\sigma/\lg\lg n)\lg n/\lg\lg n)$ & W \\
\cite{NS10} & $nH_0(S)+O(n\lg\sigma/(\eps\lg^{1-\eps} n))$ & $O((1+\frac{1}{\eps}\lg\sigma/\lg\lg n)\lg n/\lg\lg n)$ & $O((1+\frac{1}{\eps}\lg\sigma/\lg\lg n)\lg n/\lg\lg n)$ & W \\
\hline
Ours   & $nH_0(S)+O(n\lg\sigma/\lg^{1-\eps} n)$ & $O(\frac{1}{\eps^2}\lg n/\lg\lg n)$ & $O(\frac{1}{\eps^2}\lg n /\lg\lg n)$ & A \\
Ours   & $nH_0(S)+O(nH_0(S)/\lg\lg n)+O(n\lg\sigma/\lg^{1-\eps} n)$ & $O(\frac{1}{\eps^2}\lg n/\lg\lg n)$, $O(\lg n)$ for $\ra$ & $O(\lg n)$ & W \\
\end{tabular}
\end{center}
\end{sidewaystable}

Table~\ref{tab:res} gives more details on previous and our new results.
Wavelet trees can also be used to model $n \times n$ grids of points, in 
which case $\sigma=n$. Bose et al.~\cite{BHMM09} used a wavelet-tree-like
structure to solve range counting in optimal static time $O(\lg n / \lg\lg n)$,
using operations slightly more complex than rank on the wavelet tree levels.
It is conceivable that this can be turned into an $O((\lg n /\lg\lg n)^2)$
time algorithm using dynamic sequences on the wavelet tree levels. On the
other hand, $\Omega((\lg n / \lg\lg n)^2)$ is a lower bound for dynamic range
counting in two dimensions \cite{Pat07}. This seems to suggest that it is
unlikely to obtain better results than those previously known for dynamic 
wavelet trees.

In this paper we show that this dead-end can be broken by abandoning the
implicit assumption that, to provide $\acc$, $\ra$ and $\sel$ on $S$, we
{\em must} provide $\ra$ and $\sel$ on the bitmaps (or sequences over 
$[1..\rho]$). We show that
all what is needed is to {\em track} positions of $S$ downwards and upwards
along the wavelet tree. It turns out that this tracking can be done in 
{\em constant} time per level, breaking the $\Theta(\lg n / \lg\lg n)$ per-level
barrier.

As a result, we obtain the {\em optimal} time complexity
$O(\lg n/\lg\lg n)$ for all the queries (worst-case) and update operations
(amortized), independently of the alphabet size.
This is $\Theta(1+\lg\sigma/\lg\lg n)$ times faster 
than what was believed to be the ``ultimate'' solution. 
Our space is $nH_0(S) + o(n\lg\sigma)$ bits, similar to previous solutions.
We develop, alternatively, a data structure achieving worst-case time for all 
the operations, yet this raises to $O(\lg n)$ for $\ra$, insertions and 
deletions. 
\no{The set of operations can be extended to more general splitting,
concatenations, and block moves, at $O(\sigma\lg n)$ time cost per operation.
The only previous solution handling these operations \cite{NS10} achieved
$O(\sigma\lg^{1+\eps} n)$.}

Among the many applications of this result, it is worth mentioning that 
any dynamic sequence representation supporting $\ra$ and insertions in $O(t(n))$
amortized time can be used to compute the Burrows-Wheeler transform (BWT)
\cite{BW94} of a sequence $S[1,n]$ in worst-case time $O(n t(n))$. Thus our 
results allow us to build the BWT in $O(n\lg n/\lg\lg n)$ time and compressed
space. The best existing space-time tradeoffs are by Okanohara and Sadakane 
\cite{OS09}, who achieve optimal $O(n)$ time within 
$O(n \lg \sigma \lg\lg_\sigma n)$ bits, and Hon et al.~\cite{HSS09}, who 
achieve $O(n\lg\lg\sigma)$ time with $O(n\lg\sigma)$ bits. 
K\"arkk\"ainen \cite{Kar07} had obtained before $O(n\lg n + nv)$ time and
$O(n\lg n / \sqrt{v})$ extra bits for a parameter $v$.
Using less space
allows us to improve BWT-based compressors (like {\sc Bzip2}) by allowing them
to cut the sequence into larger blocks, given a fixed amount of main memory
for the compressor. 
Many other results will be mentioned in Section~\ref{sec:app}.

\section{The Wavelet Tree}
\label{sec:wavel}

Let $S$ be a string over alphabet $\Sigma=[1..\sigma]$.  We associate each
$a\in \Sigma$ to a leaf $v_a$ of a full balanced  binary tree $\mathcal{T}$. The essential idea of the wavelet tree structure \cite{GGV03} is the representation of elements from a string $S$ by bit sequences stored in
 the nodes of tree $\mathcal{T}$. We  associate a subsequence $S(v)$ 
of $S$ with every node $v$ of $\mathcal{T}$. For the root $v_r$, $S(v_r)=S$. 
In general, $S(v)$ consists of all the occurrences of symbols $a\in \Sigma_v$ 
in $S$, where $\Sigma_v = \{a \in\Sigma,~v_a~\textrm{descends from}~v\}$.
The wavelet tree does not store $S(v)$ explicitly, but just a bit vector
$B(v)$. We set $B(v)[i]=t$ if the $i$-th element
of $S(v)$ also belongs to $S(v_t)$, where $v_t$ is the $t$-th child of $v$
(the left child corresponds to $t=0$ and the right to $t=1$).
This data structure (i.e., $\mathcal{T}$ and bit vectors $B(v)$) is called a 
\emph{wavelet tree}. Since $\mathcal{T}$ has $O(\sigma)$ nodes and 
$\lceil\lg\sigma\rceil$ levels, and the bitmaps at each level add up to length
$n$, it requires $n\lceil\lg\sigma\rceil + O(\sigma\lg n)$ bits of space. If
the bitmaps $B(v)$ are compressed to $|B(v)|H_0(B(v)$ bits, the total size
adds up to $nH_0(S)+O(\sigma\lg n)$ bits \cite{GGV03}. Various surveys on
wavelet trees are available, for example \cite{NM06,Nav12}.

For any  symbol $S[i]$ and  every internal node $v$ such that $S[i]\in \Sigma_v$, 
there is exactly one bit $b_v$ in $B(v)$ that indicates in which child 
of $v$  the leaf $v_{S[i]}$ is stored. We will say that such $b_v$ 
\emph{encodes} $S[i]$ in $B(v)$; we will also say that bit $b_v$ from $B(v)$ 
\emph{corresponds} to a bit $b_u$ from $B(u)$ if both $b_v$ and $b_u$ encode 
the same symbol $S[i]$ in two nodes $v$ and $u$. 
Identifying the positions of bits that encode the same 
symbol plays a crucial role in wavelet trees. Other, more complex, 
operations rely on the ability to navigate in the tree and keep track of 
bits that encode the same symbol.

The wavelet tree encodes $S$, in the sense that it allows us to extract any
$S[i]$. To implement $\acc(S,i)$ we traverse a path from the root to the leaf 
$v_{S[i]}$. In each visited node we read the bit $b_v$ that encodes $S[i]$ and 
proceed to the corresponding bit in the $b_v$-th child of $v$. Upon arriving 
to a leaf $v_a$ we answer $\acc(S,i)=a$. 

The wavelet tree also implements operations $\ra$ and $\sel$.
To compute $\sel_a(S,i)$, we start at the position of $S(v_a)[i]$ and 
identify the corresponding bit $b_v$ in the parent $v$ of $v_a$. We continue
this process to the root until reaching a position $B(v_r)[j]$. Then the answer
is $\sel_a(S,i)=j$.
Finally, to compute $\ra_a(S,i)$, we traverse the wavelet tree from $B(v_r)[i]$
to the leaf $v_a$. At each element $b_v$ of each node $v$ in the path, 
we identify the last bit $b'$ that precedes $b_v$ and encodes an $a$. Then we
move to the bit $b_u$ corresponding to $b'$ in the $b'$-th child of $v$. Upon
arriving at position $B(v_a)[j]$, the answer is $\ra(S,i)=j$.

The standard method used in wavelet trees for identifying
 the corresponding bits 
is to maintain rank/select data structures on the bit vectors $B(v)$. 
Let $B(v)[e]=t$; we can find the offset of the 
corresponding bit in the child of $v$ by answering a query $\ra_t(B(v),e)$.
If $v$ is the $t$-th child of a node $u$, we can find the offset of the
corresponding bit in $u$ by answering a query $\sel_t(B(u),e)$. Finally, the
more complicated process of finding $b'$ needed for $\ra_a(S,i)$ is easily
solved using binary $\ra$: if $b_v$ is at position $e$ in $B(v)$, then without 
need of finding $b'$ we know that its corresponding position in the $b'$-th 
child of $v$ is $\ra_{b'}(B(v),e)$.
This approach leads to $O(\lg\sigma)$ query times in the static case because 
rank/select queries on a bit vector $B(v)$ can be answered in constant time
and $|B(v)|+o(|B(v)|)$ bits of space \cite{Mun96,Cla96}, and even using 
$|B(v)|H_0(B(v))+o(|B(v)|)$ bits \cite{RRR07}. 
However, we need $\Omega(\lg n/\lg \lg n)$ time to support rank/select and
updates on a bit vector \cite{FS89}, which multiplies the operation times in
the dynamic case. 

An improvement (for both static and dynamic wavelet trees) can be achieved by increasing the fan-out of the wavelet tree to $\rho=\Theta(\lg^{\eps} n)$ for a constant $0<\eps<1$: as before, $B(v)[e]=t$ if the $e$-th element
of $S(v)$ also belongs to $S(v_t)$ for the $t$-th child $v_t$ of $v$. 
This enables us to reduce the height of the wavelet trees and the 
query time by a $\Theta(\lg \lg n)$ factor, because the $\ra$/$\sel$ times
over alphabet $[1..\rho]$ is still constant in the static case \cite{FMMN07}
and $O(\lg n /\lg\lg n)$ in the dynamic case \cite{HM10,NS10}. 
However, it seems that further improvements that are based on dynamic rank/select queries 
in every node are not possible.

In this paper we use a different approach to identifying the 
corresponding elements. We partition sequences $B(v)$ into blocks, which are stored
in compact list structures $L(v)$. Pointers from selected positions 
in $L(v)$ to the structure $L(u)$ in children nodes $u$ (and vice versa) 
enable us to navigate between nodes of the wavelet tree in constant time. We
extend the idea to multiary wavelet trees.
While similar techniques have been used in some geometric data structures 
\cite{N11,B08}, applying them on compressed data structures 
where the bit budget is severely limited  is much more challenging.
We describe our new solution next.

\section{Basic Structure}
\label{sec:basic}

We start by describing the main components of our modified 
wavelet tree. Then, we show how our structure supports 
$\acc(S,i)$ and $\sel_a(S,i)$. In the third part of this 
section we describe additional structures that enable us to 
answer $\ra_a(S,i)$. Finally, we show how to support updates. 

\subsection{Structure}

We assume that the wavelet tree $\cT$ has node degree $\rho=\Theta(\lg^{\eps}n)$. 
We divide sequences $B(v)$ into {\em blocks} and store those blocks 
 in a doubly-linked list $L(v)$.  
Each block $G_j(v)$ contains $\Theta(\lg^3 n/\lg\rho)$ consecutive 
elements from $B(v)$, except the last, which can be smaller.
That is, $|G_j(v)|=O(\lg^3 n)$ if measured in bits.
For each  $G_j(v)$ we maintain a data structure $R_j(v)$ that supports $\acc$, $\ra$ and $\sel$ queries on elements of $G_j(v)$. Since a block contains a poly-logarithmic 
number of elements over an alphabet of size $\rho$, we can answer those queries
in $O(1)$ time using $O(|G_j(v)|/\lg^{1-\eps} n)$ additional bits
(see Section~\ref{app:A} for details).

A \emph{pointer} to an element $B(v)[e]$ consists of two parts:
a unique id of the block $G_j(v)$ that contains the offset $e$ and 
the index of $e$ in $G_j(v)$. Such a pair (block id, local index) will be
called the {\em position} of $e$ in $v$.

We maintain pointers between selected corresponding elements in $L(v)$ and its children.  
If  an element $B(v)[e]=t$ is stored in a block $G_j(v)$ and 
$B(v)[e']\not= t$ for all  $e'<e$ in $G_j(v)$ (i.e., $B(v)[e]$ is the first
occurrence of $t$ in its block), then we store 
a pointer from $e$ to the offset $e_t$ of the 
corresponding element $B(v_t)[e_t]$ in $L(v_t)$,
where $v_t$ is the $t$-th child of $v$. 
Pointers are bidirectional, that is, we also store a pointer from $e_t$ to $e$.
In addition, if $B(v)[e]$ is the first offset in its block and $B(u)[e']$ 
corresponds to $B(v)[e]$ in the parent $u$ of $v$, then we store a pointer from
$e$ to $e'$ and, by bidirectionality, from $e'$ to $e$.
All these pointers will be called 
\emph{inter-node pointers}. 
We describe how they are implemented later in this section.

It is easy to see that the number of inter-node pointers from $e$ in $L(v)$ 
to $e_t$ in $L(v_t)$, for any fixed $t$, is $\Theta(g(v))$, where $g(v)$ is the number of blocks 
in $L(v)$. Hence, the total number of pointers that point down from a node 
$v$ is bounded by $O(g(v)\rho)$. Additionally, there are $O(g(v))$ pointers 
up to the parent of $v$. Thus, the total number of pointers 
in the wavelet tree equals $O(\sum_{v\in \cT} g(v)\rho)=
O(n\lg\sigma/\lg^{3-\eps}n+\sigma\lg^\eps n)$. Note that the term
$\sigma\lg^\eps n$ is only necessary to account for nodes that have just
one block (with $o(\lg^3 n)$ bits). Since the children of those nodes
must also have just one block, we avoid storing their pointers, as we know that
all point to the same block and their index inside the block can be found with
constant-time $\ra$/$\sel$ operations inside the block from where pointers 
leave. Their upwards pointer to their parent, if the parent has more than one
block, can be represented and charged to the space of the parent. 
This yields the cleaner expression
$O(n\lg\sigma/\lg^{3-\eps}n)$ for the number of pointers.

The pointers from a block $G_j(v)$ are stored in a data structure $F_j(v)$. Using $F_j(v)$ we can find, for any 
offset $e$ in $G_j(v)$ and any $1\le t\le \rho$, the last 
$e' \le e$ in $G_j(v)$ such that there is a pointer from $e'$ to an 
offset $e'_t$ in $L(v_t)$.  
We describe in Section~\ref{app:A} how $F_j(v)$ implements the queries and 
updates in constant time.

For the root node $v_r$, we store a dynamic searchable partial-sum data 
structure $\cP(v_r)$ that contains the number of positions in each block of 
$L(v_r)$. Using $\cP(v_r)$, we can find the block $G_j(v_r)$ that contains the 
$i$-th element of $S(v_r)=S$ (query {\em search} on the partial sums), as well 
as the number of elements in all the blocks that precede a given block 
$G_j(v_r)$ (operation {\em sum} on the partial sums).
Both operations can be supported in $O(\lg n/\lg \lg n)$ time and linear space
\cite[Lem.~1]{NS10}.
The same data structures 
$\cP(v_a)$ are also stored in the leaves $v_a$ of $\cT$. 
Since $g(v_r) = O(n\lg\rho/\lg^3 n)$, and also $\sum_{a\in\Sigma} g(v_a) =
O(n\lg\rho/\lg^3 n)$, we store $O(n\lg\lg n/\lg^3 n)$ elements in the partial
sums $\cP(v_r)$ and $\cP(v_a)$, for an overall size of $O(n\lg\lg n/\lg^2 n)$
bits.

We observe that we do not store a sequence $B(v_a)$ in a leaf node
$v_a$, only in internal nodes. Nevertheless, we divide the (implicit) sequence 
$B(v_a)$ into blocks and store the  number of positions in each 
block in $\cP(v_a)$; we maintain $\cP(v_a)$ only if $L(v_a)$ consists of more than one block. Moreover we store inter-node pointers from 
the parent of $v_a$ to $v_a$ and vice versa.
Pointers in a leaf are maintained using the same rules of any other node.

For future reference, we provide the list of secondary data structures in Table~\ref{tab:not}.

\begin{table}
\caption{Structures inside any node $v$ of the wavelet tree $\cT$, 
or only in the root node $v_r$ and the leaves $v_a$. The third column gives
the extra space in bits, on top of the data, for the whole structure; here
$|data|$ is $n\lg\sigma$ in the uncompressed case and $nH_0(S)$ in the
compressed case.}
\label{tab:not}
\begin{center}
\begin{tabular}{l|l|c}
Structure & Meaning & Extra space in bits \\
\hline
$L(v)$ & List of blocks storing $B(v)$ & $O(n\lg\sigma/\lg^2 n + \sigma\lg n)$ \\
$G_j(v)$ & $j$-th block of list $L(v)$ & $O(|data|\lg\lg n/\lg n + n\lg\sigma/\lg n+\sigma\lg n)$ \\
$R_j(v)$ & Supports \ra/\sel/\acc\ inside $G_j(v)$ & $O(|data|/\lg^{1-\eps} n)$ \\
$F_j(v)$ & Pointers leaving from $G_j(v)$ & $O(n\lg\sigma/\lg^{2-\eps} n)$\\
$H_j(v)$ & Pointers arriving at $G_j(v)$ & $O(n\lg\sigma/\lg^2 n)$ \\
$P_t(v)$ & Predecessor in $L(v)$ containing symbol $t$ & $O(n\lg\sigma/\lg^{2-\eps} n)$ \\
$\cP(v)$ & Partial sums on block lengths for $v_r$ and $v_a$ & $O(n\lg\lg n/\lg^2 n)$ \\
$D_j(v)$ & Deleted elements in $G_j(v)$, for $v_r$ and $v_a$ & $O(n(\lg\lg n)^2 / \lg n)$ \\
$DEL$    & Global list of deleted elements in $\oS$ & $O(n/\lg n + n\lg\sigma/\lg^2 n)$ \\
\end{tabular}
\end{center}
\end{table}

\subsection{Access and Select Queries}

Assume the position of an element $B(v)[e]=t$ in $L(v)$ is known, 
and let $i_v$ be the index of offset $e$ in its block $G_j(v)$. 
Then the position of the corresponding offset $e_t$ in $L(v_t)$ is computed as follows.
Using $F_j(v)$, we find the index $i'_v$ of the largest $e'\le e$ in $G_j(v)$ 
such that
there is a pointer from $e'$ to some $e'_t$ in $L(v_t)$. Due to our
construction, such $e'$ must exist (it may be $e$ itself). 
Let $G_\ell(v_t)$ denote the block that contains $e'_t$, and let
$i'_t$ be the index of $e_t'$ in $G_\ell(v_t)$.
Due to our rules to define pointers, $e_t$ also 
belongs to $G_\ell(v_t)$, since if it belonged to another block $G_m(v_t)$ 
the upward pointer from $G_m(v_t)$ would point between $e'$ and $e$,
and since pointers are bidirectional, this would contradict the definition of
$e'$. 
 Furthermore, let $r_v=\ra_t(G_j(v),i_v)$ and $r'_v=\ra_t(G_j(v),i'_v)$. 
Then the index of $e_t$ is $i'_t+ (r_v-r'_v)$. 
Thus we can find the position of $e_t$ in 
$O(1)$ time if the position of $B(v)[e]=t$ is known. 

Analogously, assume we know a position $B(v_t)[e_t]$ at $G_j(v_t)$ and want to 
find the position of the corresponding offset $e$ in its parent node $v$. Using $F_j(v_t)$ 
we find the last $e'_t \le e_t$ in $G_j(v_t)$ that has a pointer to its
parent, which exists by construction (it can be the upward pointer from the 
first index in $G_j(v_t)$ or the reverse of some pointer from $v$ to $v_t$). 
Let $e'_t$ point to $e'$, with index $i_v'$ in a block $G_\ell(v)$. 
Then, by our construction, $e$ is also in $G_\ell(v)$, since if it belonged
to a different block $G_m(v)$, then the first occurrence of $t$ in $G_m(v)$
would point between $e'_t$ and $e_t$, and its bidirectional version would 
contradict the definition of $e'_t$.
Furthermore, let
$i'_t$ and $i_t$ be the indexes of $e'_t$ and $e_t$ in $G_j(v_t)$, respectively.
Then the index of $e$ is 
$\sel_t(G_\ell(v),\ra_t(G_\ell(v),i_v')+(i_t-i_t'))$.

To solve $\acc(S,i)$, we visit the nodes $v_0=v_r,v_1\ldots v_h=v_a$, where 
$h=\lg_\rho \sigma$ is the height of $\cT$,
$v_k$ is the $t_k$-th child of $v_{k-1}$ and $B(v_{k-1})[e_{k-1}]=t_k$ encodes 
$S[i]$. We do not find out the offsets $e_1,\ldots,e_h$, but just their
positions. The position of $e_0=i$ is found in $O(\lg n/\lg \lg n)$ 
time  using the partial-sums structure $\cP(v_r)$. 
If the position of  $e_{k-1}$ is known, we can find that of $e_k$ in 
$O(1)$ time, as explained above. 
When a leaf node $v_h=v_a$ is reached, we know that  $S[i]=a$.

To solve $\sel_a(S,i)$, we set $e_h=i$ and identify its position in the 
list $L(v_a)$ of the leaf $v_a$, using structure $\cP(v_a)$. 
Then we traverse the path 
$v_h,v_{h-1},\ldots,v_0=v_r$ where $v_{k-1}$ is the parent of $v_k$, until 
the root node is reached. In every node $v_k$, we find the position of
$e_{k-1}$ in $L(v_{k-1})$ that corresponds to $e_k$ as explained above. 
Finally, we compute 
the number of elements that precede $e_0$ in $L(v_r)$ using 
structure $\cP(v_r)$.  

Thus $\acc$ and $\sel$ require $O(\lg_\rho\sigma+\lg n /\lg\lg n) = 
O((\lg\sigma+\lg n)/\lg\lg n)$
worst-case time.

\subsection{Rank Queries}

We need some additional data structures for the efficient support of rank queries. 
In every node $v$ such that $L(v)$ consists of more than one block, we store a data structure $P(v)$. Using 
$P(v)$ we can find, for any $1\le t \le \rho$ and for any block 
$G_j(v)$, the last block $G_\ell(v)$  that precedes $G_j(v)$ and contains 
an element $B(v)[e]=t$. 
$P(v)$ consists of $\rho$ predecessor data structures $P_t(v)$ for 
$1\le t\le \rho$. 
We describe in 
Section~\ref{sec:rank} a way to support these predecessor queries in constant 
time in our scenario.

Let the position of offset $e$ be the $i$-th element in a block $G_j(v)$. 
$P(v)$ enables us to find the position of the last $e'\le e$ such that
$B(v)[e']=t$. First, we use $R_j(v)$ to compute
$r=\ra_t(G_j(v),i)$. If  $r>0$, then $e'$ belongs to the same 
block as $e$ and its index in the block $G_j(v)$ is
$\sel_t(G_j(v),r)$. Otherwise, we use $P_t(v)$ to find the 
last block $G_\ell(v)$ that precedes $G_j(v)$ and contains an  element 
$B(v)[e']=t$. We then find the last such element in 
$G_\ell(v)$ using $R_\ell(v)$.

Now we are ready to describe the procedure to answer $\ra_a(S,i)$. The symbol 
$a$ is represented as a concatenation of symbols 
$t_0\circ t_1\circ\ldots\circ t_h$, where each $t_k$ is between 1 and $\rho$. 
We traverse the path from the root $v_r=v_0$ to the leaf $v_a=v_h$.  
We find the position of $e_0=i$ in $v_r$ using 
the data structure $\cP(v_r)$. In each node $v_k$, $0\le k< h$, 
 we identify the position of the last element $B(v_k)[e'_k]=t_k$ that precedes 
$e_k$, using $P_{t_k}(v_k)$. 
Then we find the offset $e_{k+1}$ in the list $L(v_{k+1})$ that 
corresponds to $e'_{k}$.

When our procedure reaches the leaf node $v_h$, the element $B(v_h)[e_h]$ encodes the last 
symbol $a$ that precedes $S[i]$. We know the position of offset $e_h$, say
index $i_h$ in its block $G_\ell(v_h)$. Then we find the number $r$ of 
elements in all the blocks that precede $G_\ell(v_h)$ using $\cP(v_h)$. 
Finally, $\ra_a(S,i)=r+i_h$.

Since structures $P_t$ answer queries in constant time, the overall time for
$\ra$ is $O(\lg_\rho\sigma+\lg n/\lg\lg n) = O((\lg\sigma + \lg n)/ \lg\lg n)$.

\subsection{Updates} \label{sec:H}

Now we describe how inter-node pointers are implemented.
We say that an element of $L(u)$ is \emph{pointed} if there 
is a pointer to its offset. 
Unfortunately, we cannot store the local index of a pointed 
element in the pointer: when a new element is inserted into 
a block, the indexes of all the elements that follow it are incremented 
by $1$. Since a block can contain $\Theta(\lg^3 n/\lg\rho)$ pointed 
elements, we would have to update that many
pointers after each insertion and deletion. 

Therefore we resort to the following two-level scheme. 
Each pointed element in a block is assigned a unique id. 
When a new element is inserted, we assign it the id 
$max_-id+1$, where $max_-id$ is the maximum 
id value used so far. 
We also maintain a data structure $H_j(v)$ for each block 
$G_j(v)$ that enables us to find the position 
of a pointed element if its id in $G_j(v)$ is known.
Implementation of $H_j(v)$ is based on standard 
word RAM techniques and a table that contains ids of the 
pointed elements; details are given in Section~\ref{app:A}.

We describe now how to insert a new symbol $a$ into $S$ at position $i$. 
Let $e_0,e_1,\ldots,e_h$ be the offsets of the elements that will encode 
$a = t_0 \circ \ldots \circ t_h$ in $v_r=v_0,v_1,\ldots,v_h=v_a$. 
We can find the position of $e_0=i$ in $L(v_r)$ in $O(\lg n/\lg \lg n)$ time
using $\cP(v_r)$, and insert $t_0$ at that position, $B(v_r)[e_0]=t_0$.
Now, given the position of $e_k$, in $L(v_k)$, where $B(v_k)[e_k]=t_k$ has
just been inserted,
we find the position of the last $e'_k<e_k$ such that 
$B(v_k)[e'_k]=t_k$, in the same way as  for rank queries.
Once we know the position of $e'_k$ in $L(v_k)$, we find 
the position of $e''_{k+1}$ in $L(v_{k+1})$ that corresponds to $e'_k$. 
The element $t_{k+1}$ must then be inserted into $L(v_{k+1})$ 
immediately after $e''_{k+1}$, at position $e_{k+1} = e''_{k+1}+1$.

The insertion of a new element $B(v_k)[e_k]=t$ into a block $G_j(v_k)$ 
is handled by structure $R_j(v_k)$ and the memory manager of the block.
We must also update structures
$F_j(v_k)$ and $H_j(v_k)$ to keep the correct alignments, and possibly
to create and destroy a constant number inter-node pointers to maintain 
our invariants.
Also, since pointers are bidirectional, a constant number of inter-node 
pointers in the parent and children of node $v_k$ may be updated. All those 
changes can be done in $O(1)$ time; see Section~\ref{app:A} for the details. 
Insertions may also require updating structures $P_t(v_k)$, which require
$O(1)$ amortized time, see Section~\ref{sec:rank}. 
Finally, if $v_k$ is the root node or a leaf, we also update $\cP(v_k)$.
This update is only by $\pm 1$, so it requires just
$O(\lg n /\lg\lg n)$ time \cite[Lem.~1]{NS10}.

If the number of elements in $G_j(v_k)$ exceeds 
$2\lg^3 n$, we split $G_j(v_k)$ evenly into two blocks, $G_{j_1}(v_k)$ and 
$G_{j_2}(v_k)$.
Then, we rebuild the data structures $R$, $F$ and $H$ for the two new blocks.
Note that there are inter-node pointers to $G_j(v_k)$ that now could
become dangling pointers, but all those can be known from $F_j(v_k)$, since
pointers are bidirectional, and updated to point to the right places in
$G_{j_1}(v_k)$ or $G_{j_2}(v_k)$.
Finally, if $v_k$ is the root or a leaf, then $\cP(v_k)$ is updated,
meaning that we replace an existing element by two.

The total cost of splitting a block is dominated by that
of building the new data structures $R$, $F$ and $H$.
These are easily built in $O(\lg^3 n/\lg\rho)$ time. Since
we split a block $G_j(v)$ at most once per sequence of $\Theta(\lg^3 n)$ 
insertions in $G_j(v)$, the amortized cost incurred by splitting a block 
is $o(1)$. Therefore the total cost of an insertion in $L(v)$ is $O(1)$.
The insertion of a new 
symbol leads to $O(\lg_\rho \sigma)$ insertions into 
lists $L(v)$. 

Our partial-sums structures $\cP(v_r)$ and $\cP(v_a)$ do not support updates with large 
values. Inserting a new value for $G_{j_2}(v_k)$ and moving part of the value 
of $G_{j}(v_k)$ to $G_{j_2}(v_k)$ can be done in 
$O(\lg^3 n/\lg \lg n)$ time by subtracting $O(\lg n)$ units from
the value for $G_j(v)$ until it becomes $|G_{j_1}(v_k)|$, then inserting a number after
it with value zero and increasing it by $O(\lg n)$ units until it becomes
$|G_{j_2}(v_k)|$. Each such increment/decrement and insertion takes
$O(\lg n /\lg\lg n)$ time \cite[Lem.~1]{NS10} and we carry it out
$O(\lg^2 n/\lg\rho)$ times. Still this total cost amortizes to $o(1)$ per operation.

Hence, the total cost of an insertion is $O(\lg_\rho\sigma+\lg n/\lg\lg n) =
O((\lg \sigma+ \lg n)/\lg\lg n)$.

We describe next how deletions are handled, where
we also describe the data structure $P(v)$. 

\section{Lazy Deletions and Data Structure $P(v)$}
\label{sec:rank}

We do not process deletions immediately, but in lazy form: we do not
maintain exactly $S$ but a supersequence $\oS$ of it. When a symbol $S[i]=a$ is 
deleted from $S$, we retain it in $\oS$ but take a notice that 
$S[i]=a$ is deleted. When the number of deleted symbols exceeds 
a certain threshold, we expunge from the data structure all the elements 
marked as deleted.
We define $\oB(v)$ and the list $\oL(v)$ for the sequence 
$\oS$ in the same way as $B(v)$ and $L(v)$ are defined for 
$S$.

Since elements of $\oL(v)$ are never removed, we can implement 
$P(v)$ as an insertion-only data structure. For any 
$t$, $1\le t \le \rho$, we store information about all the blocks 
of a node $v$ in a data structure $P_t(v)$. 
$P_t(v)$ contains one element for each block $G_j(v)$ and is 
implemented as an incremental split-find data structure that supports 
insertions and splitting in $O(1)$ amortized time and queries in $O(1)$ 
worst-case time 
\cite{IA84}. The splitting positions in $P_t(v)$ are the blocks $G_j(v)$ that 
contain an occurrence of $t$, so the operation ``find'' in $P_t(v)$ allows us 
to locate, for any $G_j(v)$, the last block preceding $G_j(v)$ that contains 
an occurrence of $t$. 

The insertion of a symbol $t$ in $\oL(v)$ may induce a new split in $P_t(v)$. 
Furthermore, overflows in a block $G_j(v)$, which convert it into two blocks 
$G_{j_1}(v)$ and $G_{j_2}(v)$, induce insertions in $P_t(v)$.
Note that an overflow in $G_j(v)$ triggers $\rho$ insertions in the $P_t(v)$
structures, but this $O(\rho)$ time amortizes to $o(1)$ because overflows
occur every $\Theta(\lg^3 n/\lg\rho)$ operations.

Structures $P_t(v)$ do not support ``unsplitting'' nor removals.
The replacement of $G_j(v)$ by  $G_{j_1}(v)$ and $G_{j_2}(v)$ is implemented as 
leaving in 
$P_t(v)$ the element corresponding to $G_j(v)$ and inserting one corresponding 
to either $G_{j_1}(v)$ or $G_{j_2}(v)$. If $G_j(v)$ contained $t$, then at
least one of $G_{j_1}(v)$ and $G_{j_2}(v)$ contain $t$, and the other can be
inserted as a new element (plus possibly a split, if it also contains $t$).

We need some additional data structures to support lazy deletions. 
A data structure $\oP(v)$ stores the number of non-deleted elements 
in each block of $\oL(v)$ and supports partial-sum queries. 
We will maintain $\oP(v)$ in the root of the wavelet tree and 
in all leaf nodes. Moreover, we maintain a data structure 
$D_j(v)$ for every block $G_j(v)$, where $v$ is either the root 
or a leaf node. $D_j(v)$ can be used to count the number of deleted 
and non-deleted elements before the $i$-th element in a block $G_j(v)$ for any query index $i$, as well as to find the index in $G_j(v)$ of the 
$i$-th non-deleted element. 
The implementation of $D_j(v)$ is described in Section \ref{app:A}. 
We can use $\oP(v)$ and $D_j(v)$ to find the index $\oi$ in $\oL(v)$
where the $i$-th non-deleted element occurs, and to count the number 
of non-deleted elements that occur before the index $\oi$ in $\oL(v)$. 

We also store a global list {\em DEL} that contains, in any order, all 
the deleted symbols that have not yet been expunged from the wavelet tree. 
For any symbol $\oS[i]$ in the list {\em DEL} we store a pointer to the 
offset $e$ in $\oL(v_r)$ that encodes $\oS[i]$. Pointers in 
list {\em DEL} are implemented in the same way  as inter-node
 pointers.

\subsection{Queries}

Queries are answered very similarly to 
Section~\ref{sec:basic}. The main idea is that we can essentially ignore
deleted elements except at the root and at the leaves.
\begin{description}
\item[$\acc(S,i)$:] 
Exactly as in Section 3, except that $e_0$ encodes the $i$-th 
non-deleted element in $\oL(v_r)$, and is found using $\oP(v_r)$ and $D_j(v_r)$.

\item[$\sel_a(S,i)$:] We find the position of the offset $e_h$ of the 
$i$-th non-deleted element in $\oL(v_h)$, where $v_h=v_a$, using $\oP(v_a)$ and some $D_j(v_a)$. 
Then we move up in the tree exactly as in Section~\ref{sec:basic}.
When the root node $v_0=v_r$ is reached, we count the number of non-deleted 
elements that precede offset $e_0$ using $\oP(v_r)$. 

\item[$\ra_a(S,i)$:] We find the position of the offset $e_0$ of the $i$-th 
non-deleted element in $\oL(v_r)$. Let $v_k,t_k$ be defined as in Section~\ref{sec:basic}. 
In every node $v_k$, we find the last offset $e'_k \le e_k$ such that
$\oB(v_k)[e'_k]=t_k$. Note that this element may be a deleted one, but it still drives us to
the correct position in $\oL(v_{k+1})$. We proceed exactly as in 
Section~\ref{sec:basic} until we arrive at a leaf $v_h=v_a$.
At this point, we count the number of non-deleted elements that precede 
offset $e_h$ using $\oP(v_a)$ and $D_j(v_a)$.
\end{description}

\subsection{Updates} \label{sec:rankdel}

Insertions are carried out just as in Section~\ref{sec:basic}. 
The only difference is that we also update the data structure $D_j(v_k)$ when 
an element $B(v_k)[e_k]$ that encodes the inserted symbol $a$ is added to a block 
$G_j(v_k)$. When a symbol $S[i]=a$ is deleted, we append it to the list {\em DEL} of
deleted symbols. Then we visit each block $G_j(v_k)$ containing the element 
$B(v_k)[e_k]$ that encodes $S[i]$ and update the data structures $D_j(v_k)$. 
Finally, $\oP(v_r)$ and $\oP(v_a)$ are also updated. This takes in total
$O(\lg_\rho\sigma+\lg n/\lg\lg n)$ time.

When the number of symbols in the list {\em DEL} reaches
$n/\lg^2 n$, we perform a \emph{cleaning} procedure 
and get rid of all the deleted elements. Therefore {\em DEL} never requires
more than $O(n/\lg n)$ bits, and the overhead due to storing deleted symbols
is $O(n\lg\sigma/\lg^2 n)$ bits.

Let $B(v_k)[e_k]$, $0\le k\le h$, be the sequence of elements that encode 
a symbol $\oS[i]\in DEL$. The method for tracking the elements 
$B(v_k)[e_k]$, removing them from their blocks $G_j(v_k)$, 
and updating the block structures, is symmetric to the insertion procedure 
described in Section~\ref{sec:basic}.  
In this case we do not need the predecessor
queries to track the symbol to delete, as the procedure is similar to that
for accessing $S[i]$. When the size of a block $G_j(v_k)$ falls below
$(\lg^3 n)/2$ and it is not the last block of $L(v_k)$, we merge it with 
$G_{j+1}(v_k)$, and then split the result if its size exceeds $2\lg^3 n$.
This retains $O(1)$ amortized time per deletion in any node $v_k$, 
including the updates to $\cP(v_k)$ structures, and this adds up to
$O((\lg\sigma+\lg n)/\lg\lg n)$ amortized time per deleted symbol.

Once all the pointers in {\em DEL} are processed, we rebuild from scratch 
the structures $P(v)$ for all nodes $v$. The total size of all the
$P(v)$ structures is $O(\rho n\lg\sigma/\lg^3 n)$ elements. 
Since a data structure 
for incremental split-find is constructed in linear time, 
all the $P(v)$s are rebuilt in $O(n\lg\sigma/\lg^{3-\eps}n)$ time. 
Hence the amortized time to rebuild the $P(v)$s is 
$O(\lg\sigma/\lg^{1-\eps}n)$, which does not affect the amortized time
$O((\lg\sigma+\lg n) / \lg\lg n)$ to carry out the effective deletions.

\section{Changes in $\lg n$ and Alphabet Independence}  \label{sec:space}

Note that our structures depend on the value of $w=\lg n$, so they
should be rebuilt when $\lg n$ changes. We use $w=\lceil \lg n\rceil$ as a
fixed value and rebuild the structure from scratch when $n$ reaches another 
power of two (more precisely, we use words of $w=\lceil \lg n\rceil$ bits
until $\lceil \lg n\rceil$ increases by 1 or decreases by 2, and only then
update $w$ and rebuild). These reconstructions do not affect the amortized 
complexities, and the slightly larger words waste an $O(1/\lg n)$ extra space 
factor in the redundancy.

We take advantage of using a fixed $w$ value 
to get rid of the alphabet dependence. If
$\lg\sigma \le w$, our time complexities are the optimal 
$O(\lg n /\lg\lg n)$. However, if $\sigma$ is larger, this means that not all
the alphabet symbols can appear in the current sequence (which contains at most
$n \le 2^w < \sigma$ distinct symbols). Therefore, in this case we create
the wavelet tree for an alphabet of size $s=2^w$, not $\sigma$
(this wavelet tree is created when $w$ changes).
We also set up a {\em mapping}
array $SN[1,\sigma]$ that will tell to which value in $[1..s]$ is a symbol 
mapped, and
a {\em reverse mapping} $NS[1,s]$ that tells to which original symbol in
$[1..\sigma]$ does a mapped symbol correspond. Both $SN$ and $NS$ are initialized
in constant time \cite[Section III.8.1]{Meh84} and require $O(\sigma\lg n +
n\lg\sigma)$ bits of space. Since this is used only when $\sigma > n$, the
space is $O(\sigma\lg n)$.

Upon operations $\ra_a(S,i)$ and $\sel_a(S,j)$, the symbol $a$ is mapped using
$SN$ (the answer is obvious if $a$ does not appear in $SN$) in constant time.
The answer of operation $\acc(S,i)$ is mapped using $NS$ in constant time as 
well. Upon insertion of $a$, we also map $a$ using $SN$. If not present in $SN$,
we find a free slot $NS[i]$ (we maintain a list of free slots) and assign 
$NS[i] = a$ and $SN[a]=i$.
When the last occurrence of a symbol $a$ is deleted we return its slot to
the free list and unitialize its entry in $SN$. In this way, when $\lg\sigma
> \lg n$, we can support all the operations in time $O(\lg s / \lg\lg s) =
O(\lg n / \lg\lg n)$.

We are ready to state a first version of our result, not yet compressing
the sequence. In Section~\ref{app:A} it is seen
that the time for the operations is $O(1/\eps)$. Since the
height of the wavelet tree is $\lg_\rho \min(\sigma,s) = O(\frac{1}{\eps}\lg n /
\lg\lg n)$, then we have $O(\frac{1}{\eps^2}\lg n/\lg\lg n)$ time 
for all the operations.

As for the space,
we show in Section~\ref{app:A} how to manage the data in blocks $G_j(v)$ 
so that all the elements stored in lists $L(v)$ use $n\lg \sigma$ bits, plus
the overhead $O(n\lg\sigma\lg\lg n/\lg n + \sigma\lg n)$ of the data
organization and the memory manager.
The internal structures
$R_j(v)$ add up to $O(n\lg\sigma/\lg^{1-\eps} n)$ extra bits.
Since there are $O(n\lg\sigma/\lg^3 n + \sigma)$ blocks overall,
all the pointers between blocks of 
the same lists add up to $O(n\lg\sigma/\lg^2 n + \sigma\lg n)$ bits. 
All the data structures $\cP(v)$ add up to $O(n\lg\rho/\lg^2 n)$ bits. 
We have shown that there are
$O(n\lg\sigma/\lg^{3-\eps} n)$ inter-node pointers, hence all inter-node 
pointers (i.e., $F_j$ and $H_j$ structures) use 
$O(n\lg\sigma/\lg^{2-\eps}n)$ bits. 
Structures $P_t(v)$ use $O(n\lg\sigma/\lg^{2-\eps} n)$
bits as they have $\rho$ integers per block, and $DEL$ takes $O(n/\lg n)$ bits
plus the overhead of $O(n\lg\sigma/\lg^2 n)$ of keeping deleted elements.
The overall space is then 
$n\lg\sigma + O(n\lg\sigma/\lg^{1-\eps} n) + O(\sigma\lg n)$ bits.
(Note that when $\sigma > n$ we use an alphabet of size $O(n)$, but then still
we need the $SN$ mapping, that takes $O(\sigma\lg n)$ bits.)
This gives our first result.

\begin{theorem}
\label{thm:optimal}
A dynamic string $S[1,n]$ over alphabet $[1..\sigma]$ 
can be stored in a structure using 
$n\lg\sigma + O(n\lg\sigma/\lg^{1-\eps} n) + O(\sigma\lg n)$ bits,
for any $0<\eps<1$,
and supporting queries
$\acc$, $\ra$ and $\sel$ in time $O(\frac{1}{\eps^2}\lg n/\lg \lg n)$.
Insertions and deletions of symbols 
are supported in $O(\frac{1}{\eps^2}\lg n/\lg \lg n)$ amortized time. 
\end{theorem}

\section{Compressed Space}
\label{sec:rrr}

We now compress the space of the data structure to
zero-order entropy ($nH_0(S)$ plus redundancy).
We show how a different encoding of the bits within the blocks reduces
the $n\lg\sigma$ to $nH_0(S)$ in the space without affecting the time
complexities. 

Raman et al.\ \cite{RRR07} describe an encoding for a bitmap $B[1,n]$ that
occupies $nH_0(B) + O(n\lg\lg n /\lg n)$ bits of space. It consists of
cutting the bitmap into chunks of length $b = (\lg n)/2$ and encoding each
chunk $i$ as a pair $(c_i,o_i)$: $c_i$ is the {\em class}, which indicates how 
many 1s are there in the chunk, and $o_i$ is the {\em offset}, which is the 
index of this particular chunk within its class. The $c_i$ components add up to
$O(n\lg\lg n / \lg n)$ bits, whereas the $o_i$ components add up to $nH_0(B)$.
Navarro and Sadakane \cite[Sec.\ 8]{NS10} describe a technique to maintain a 
dynamic bitmap in this format. They allow the chunk length $b$ to vary, so 
they encode triples $(b_i,c_i,o_i)$ maintaining the invariant that 
$b_i+b_{i+1} > b$ for any $i$. They show that this retains the same space, 
and that each update affects $O(1)$ chunks.

We extend this encoding to handle an alphabet $[1..\rho]$ 
\cite{FMMN07}, so that $b = (\lg_\rho n)/2$ symbols, and each chunk is
encoded as a tuple $(b_i,c_i^1,\ldots,c_i^\rho,o_i)$ where $c_i^t$ counts
the occurrences of $t$ in the block. The classes
$(b_i,c_i^1,\ldots,c_i^\rho)$ use $O(\rho n \lg\lg n / \lg n)$ bits, and
the offsets still add up to $nH_0(B)$. Blocks are encoded/decoded in
$O(1)$ time, as the class takes $O(\rho \lg\lg n) = o(\lg n)$ bits and
the block encoding requires at most $O(\lg n)$ bits.
In Section~\ref{app:A} we show how using compressed chunks does not affect
their handling inside blocks.

The sum of the local entropies of the chunks, across the whole 
$L(v)$, adds up to $nH_0(B_v)$, and these add up to $nH_0(S)$ \cite{GGV03}.
The redundancy over the entropy is $O(\rho \lg\lg n)$ bits per
miniblock, adding up to $O(n H_0(S) \lg\lg n / \lg^{1-\eps} n)$ bits, and
we have also a fixed redundancy of $O(n\lg\sigma/\lg n + n(\lg\lg n)^2/\lg n+
\sigma\lg n)$, according to
Section~\ref{app:A}.
The fact that we store $\oS$ instead of $S$, with up to $O(n/\lg^2 n)$ spurious
symbols, can increase $nH_0(S)$ up to $nH_0(\oS) \le nH_0(S)+O(n/\lg n)$ bits.
Thus we get the following result, for any desired $0<\eps<1$.

\begin{theorem}
\label{thm:compr}
A dynamic string $S[1,n]$ over alphabet $[1..\sigma]$ 
can be stored in a structure using 
$nH_0(S) + O(nH_0(S)/\lg^{1-\eps}n+n\lg\sigma/\lg n + n(\lg\lg n)^2/\lg n + \sigma\lg n) = nH_0(S) + O(n\lg\sigma/\lg^{1-\eps} n + \sigma\lg n)$ bits,
for any $0<\eps<1$,
and supporting queries
$\acc$, $\ra$ and $\sel$ in time $O(\frac{1}{\eps^2}\lg n/\lg \lg n)$.
Insertions and deletions of symbols 
are supported in $O(\frac{1}{\eps^2}\lg n/\lg \lg n)$ amortized time. 
\end{theorem}

\section{Worst-Case Complexities} \label{sec:worst}

While in previous sections we have obtained optimal time and compressed space,
the time for the update operations is amortized. In this section we derive
worst-case time complexities, at the price of losing the time optimality,
which will now become logarithmic for some operations. 
Along the rest of the section we remove the various sources of amortization 
in our solution.

\subsection{Block Splits and Merges} \label{sec:indels}

Our amortized solution splits overflowing blocks and rebuilds the two new
blocks from scratch (Section~\ref{sec:H}). Similarly, it merges underflowing 
blocks (as a part of the cleaning of the global $DEL$ list in 
Section~\ref{sec:rankdel}). This gives good amortized times but in the worst
case the cost is $\Omega(\lg^3 n/\lg\lg n)$.

We use a technique \cite{GN09} that avoids global rebuildings. A block
is called {\em dense} if it contains at least $\lg^3 n$ bits, and {\em
sparse} otherwise. While sparse blocks of any size (larger than zero) are 
allowed, we maintain the invariant that no two consecutive sparse blocks may 
exist. This retains the fact that there are $O(n\lg\sigma/\lg^3 n + \sigma)$ 
blocks in the data structure. The maximum size of a block will be $2\lg^3 n$
bits.
When a block overflows due to an insertion, we move its last element to the
beginning of the next block. If the next block would also overflow,
then we are entitled to create a new sparse block between both dense blocks, 
containing only that element. Analogously, when a deletion converts a dense
block into sparse (i.e., it falls below length $\lg^3 n$), we check if both
neighbors are dense. If they are, the current block can become sparse. If,
instead, there is a sparse neighbor, we move its first/last element
into the current block to avoid it becoming sparse. If this makes that sparse
neighbor block become of size zero, we remove it.

Therefore, we only create and destroy empty blocks, and move a constant 
number of elements to neighboring blocks. This can be done in constant
worst-case time. It also simplifies the operations on the partial-sum
data structures $\cP(v)$, since now only updates by $\pm 1$ and insertions/deletions
of elements with value zero are necessary, and these are carried out in
$O(\lg n / \lg\lg n)$ worst case time \cite[Lem.~1]{NS10}.
Recall that $\lg n$ is fixed in each instance of our data
structure, so the definition of sparse and dense is static.

\subsection{Split-Find Data Structure and Lazy Deletions}

The split-find data structure \cite{IA84} we used in Section~\ref{sec:rank}
to implement the $P_t$ structures has constant amortized insertion time. We
replace it by  another one \cite[Thm 4.1]{Mor03} achieving $O(\lg\lg n)$
worst-case time. Their structure handles a list of colored elements (list nodes), where
each element can have $O(1)$ colors (each color is a positive integer bounded by $O(\log^{\eps} n)$ for a constant $0<\eps<1$). 
We will only use list nodes with 0 or 1 color.
The operations of interest to us are: creating a new list node without colors, 
assigning or removing a color to/from a list node, and finding the last list node 
preceding a given node and having some given color. Node deletions are not
supported. The number of list nodes must be smaller than a certain upper bound $n'$,
and the operations cost $O(\lg\lg n')$. In our case, since $\lg n$ is fixed,
we can use $n'=2^w=O(n)$ as the upper bound.

We use $\rho$ colors, one per symbol in the sequences. Each time we create
a block, we add a new uncolored node to the list, with a 
bidirectional pointer to the block. Each time we insert a symbol
$t \in [1..\rho]$ for the first time in a block, we add a new node colored
$t$ to the list, right after the uncolored element that represents the block,
and also set a bidirectional pointer between this node and the block.

We cannot use the lazy deletions mechanism of Section~\ref{sec:rank}, as it
gives only good amortized complexity. We carry out the deletions immediately
in the blocks, as said in Section~\ref{sec:indels}.
Each time the last occurrence of a symbol $t \in [1..\rho]$ is deleted from a
block, we remove the color from the corresponding list node (if the symbol reappears
later, we reuse the same node and color it, instead of creating a new one). 

Therefore, finding
the last block where a symbol $t$ appears, as needed by the $\ra$ query and
for insertions, corresponds to finding the last list node colored $t$ and
preceding the uncolored node that represents the current block.

Since list nodes cannot be deleted, when a block disappears its (uncolored)
list nodes are left without an associated block. This does not alter the result
of queries, but there is the risk of maintaining too many
useless nodes. We permanently run an incremental list
``copying'' process, traversing the current list of blocks and inserting the
corresponding nodes into a new list. This new list is also updated, together
with the current list, on operations concerning the blocks already copied.
When the new list is ready it becomes
the current list and the previous list is incrementally deleted. In
$O(n\rho\lg\rho/\lg^3 n)$ steps we have copied the current list; by this time the
number of useless nodes is at most $O(n\rho\lg\rho/\lg^3 n)$ and just poses 
$O(n\lg\lg n/\lg^{2-\eps}n)$ bits of space overhead.

Note that blocks must manage the sets of up to $\rho$ pointers to their
colored nodes. This is easily handled in constant time with the same
techniques used for structure $F_j(v)$ in Section~\ref{app:A}.

Since the colored list data structure requires $O(\lg\lg n)$ time,
operations $\ra$ and insert take worst-case time $O(\frac{1}{\eps}\lg n)$, 
whereas access, select and delete still stay in $O(\frac{1}{\eps^2}\lg n / \lg\lg n)$.

\subsection{Changes in $\lg n$}

As an alternative to reconstructing the whole structure when $n$ doubles
or halves, M\"akinen and
Navarro \cite{MN08} describe a way to handle this problem without affecting
the space nor the time complexities, in a worst-case scenario.
The sequence is cut into a prefix, a middle part, and a suffix. The middle
part uses a fixed value $\lceil \lg n\rceil$, the prefix uses $\lceil \lg n
\rceil-1$ and the suffix uses $\lceil \lg n\rceil+1$. Insertions and deletions
trigger slight expansions and contractions in this separation, so that when
$n$ doubles all the sequence is in the suffix part, and when $n$ halves all
the sequence is in the prefix part, and we smoothly move to a new value of
$\lg n$. This means that the value of $\lg n$ is fixed for any instance of our
data structure. Operations access, rank and select, as well as insertions
and deletions, are easily adapted to handle this split string.

Actually, to have sufficient time to build universal tables of size 
$O(n^\alpha)$ for $0<\alpha<1$, the solution \cite{MN08} maintains the sequence
split into five, not three, parts. This gives also sufficient time to build
any universal table we need to handle block operations in constant time, as
well as to build the wavelet tree structures of the new partitions.

\subsection{Memory Management Inside Blocks}

The EAs of Lemma~\ref{lem:EA} (Section~\ref{app:A})
have amortized times to grow and shrink.
Converting those to worst-case time requires a constant space overhead
factor. While this is acceptable for the EAs of structures $Tbl$ in 
Section~\ref{app:A}, they
raise the overall space to $O(nH_0(S))$ bits if used to maintain the main data.
Instead, we get rid completely of the EA mechanism to maintain the data, and 
use a single large memory area for all the miniblocks
of Section~\ref{app:A}, using Munro's technique \cite{Mun86}. 

The problem of using a single memory area is that the pointers to the
miniblocks require $\Theta(\lg n)$ bits, which is excessive because miniblocks
are also of $\Theta(\lg n)$ bits. Instead, we use slightly larger miniblocks,
of $\Theta(\lg n \lg\lg n)$ bits. This makes the overhead due to pointers
to miniblocks $O(|G_j(v)|/\lg\lg n)$, adding up to additional
$O(nH_0(S)/\lg\lg n + n\lg\sigma/\lg^{1-\eps} n)= o(n\lg\sigma)$ bits.

The price of using larger miniblocks is that now the operations on blocks
are not anymore constant time because they need to traverse a miniblock,
which takes time $O(\lg\lg n)$. We can still retain constant time for the
query operations, by considering {\em logical} miniblocks of $\Theta(\lg n)$
bits, which are stored in {\em physical} areas of $\Theta(\lg\lg n)$
miniblocks. However, update operations like insert and delete must shift 
all the data in the miniblock area and possibly relocate it in the memory 
manager, plus updating pointers to all the logical miniblocks displaced or 
relocated. This costs $O(\lg\lg n)$ time per insertion and deletion.
This completes our result.

\begin{theorem}
\label{thm:worstcase}
A dynamic string $S[1,n]$ over alphabet $[1..\sigma]$ 
can be stored in a structure using 
$nH_0(S) + O(nH_0(S)/\lg\lg n) + O(n\lg\sigma/\lg^{1-\eps} n)+O(\sigma \lg n)
= nH_0(S)+o(n\lg\sigma)+O(\sigma\lg n)$ bits, for any constant $0<\eps<1$,
and supporting queries $\acc$ and $\sel$ in worst-case time
$O(\frac{1}{\eps^2}\lg n /\lg\lg n)$, and query $\ra$, insertions and deletions in  worst-case time $O(\frac{1}{\eps}\lg n)$.
\end{theorem}

\no{
\section{Extensions}
\label{sec:ext}

\subsection{Handling General Alphabets}
\label{sec:alphabet}
}

\section{Data Structures for Handling Blocks} \label{app:A}

We describe the way the data is stored in blocks $G_j(v)$, as well as 
the way the various structures inside blocks operate. All the data 
structures are based on the same idea: We maintain a tree with node degree 
$\lg^{\delta}n$ and leaves that contain $O(\lg n)$ bits. Since elements 
within a block can be addressed with $O(\lg \lg n)$ bits, each internal node 
and each leaf fits into one machine word. Moreover, we can support searching 
and basic operations  in each node in constant time.  

\subsection{Data Organization}

The block data is physically stored as a sequence of {\em miniblocks} of
$\Theta(\lg_\rho n)$ symbols, or $\Theta(\lg n)$ bits. Thus there are 
$O(\lg^2 n)$ mini\-blocks in a block. These miniblocks will be the 
leaves of a $\tau$-ary tree $T$, for $\tau = \Theta(\lg^\delta n)$ and some 
constant $0<\delta<1$. The height of this tree is constant, $O(1/\delta)$. 
Each node of $T$ stores $\tau$ counters telling the number 
of symbols stored at the leaves that descend from each child. This requires 
just $O(\tau \lg\lg n) = o(\lg n)$ bits. To access any position of $G_j(v)$,
we descend in $T$, using the counters to determine the correct child.
When we arrive at a leaf, we know the local offset of the desired symbol
within the leaf, and can access it directly. Since the counters fit in 
less than a machine word, a small universal table gives the correct child in 
constant time, therefore we have $O(1)$ time access to any symbol (actually to any 
$\Theta(\lg_\rho n)$ consecutive symbols).

Upon insertions or deletions, we arrive at the correct leaf, insert or delete 
the symbol (in constant time because the leaf contains $\Theta(\lg n)$ bits
overall), and update the counters in the path from the root (in constant time
as they have $o(\lg n)$ bits). The leaves may have $\lg n$ to $2\lg n$ bits.
Splits/merges upon overflows/underflows are handled as usual, and can be
solved in a constant number of $O(1)$-time operations ($T$
operates as a B-tree; internal nodes may have $\tau$ to $2\tau$ children).

The space overhead due to the nodes of $T$ is
$O(|G_j(v)| \lg\lg n / \lg n)$ bits, where we measure
$|G_j(v)|$ in bits, not symbols. The factor $\tau$ disappears 
because each leaf of $T$ has $\tau$ miniblocks. 

We consider now the space used by the data itself.
In order not to waste space, the miniblock leaves are stored using a memory
management structure by Munro \cite{Mun86}. For our case, it allows us to 
allocate, free, and access  miniblocks of length up to $2\lg n$ in
constant time. Its space waste, given that our pointers are internal to blocks
and require $O(\lg\lg n)$
bits, is $O(\lg\lg n)$ per allocated miniblock, which adds up to
$O(|G_j(v)|\lg\lg n / \lg n)$, plus a global redundancy
of $O(\lg^2 n)$ bits.
If we used one allocation structure per block, handling its miniblocks,
the global redundancy of $O(\lg^2 n)$ bits per block would add
$O(n\lg\sigma/\lg n + \sigma\lg^2 n)$ bits 
overall. This is reduced to $O(n\lg\sigma/\lg^2 n + \sigma\lg n)$ by
using one allocation structure per group of $\lg n$ blocks. This
reduces the overhead of the structures and the address space is
still of size $O(\lg^4 n)$, so pointers can still be of length $O(\lg\lg n)$.

Each allocation structure uses a memory area of fixed-size cells (inside which 
the variable-length miniblocks are stored) that grows or shrinks at the end
as miniblocks are created or destroyed. A structure to store those memory
areas with fixed-size cells and allowing them to grow and shrink
is the {\em extendible array (EA)} \cite{RR03}. We need to handle a
set of $O(n\lg\sigma/\lg^4 n+\sigma/\lg n)$ EAs, what is called a 
{\em collection of extendible 
arrays}. It supports accessing any cell of any EA, letting
any EA grow or shrink by one cell, and create and destroy EAs. 
The following lemma,
simplified from the original \cite[Lemma 1]{RR03}, and using words of $\lg n$
bits, is useful.

\begin{lemma} \label{lem:EA}
A collection of $a$ EAs of total size $s$ bits can be represented using
$s + O(a \lg n + \sqrt{sa\lg n})$ bits of space, so that the operations of 
creation of an empty EA and access take constant worst-case time, whereas 
grow/shrink take constant amortized time. An EA of $s'$ bits can be destroyed 
in time $O(s'/\lg n)$.
\end{lemma}

In our case $a = O(n\lg\sigma/\lg^4 n+\sigma/\lg n)$ and $s=O(n\lg\sigma)$, so 
the space overhead posed by the EAs is 
$O(n\lg\sigma/\lg^3 n + \sigma +
   n\lg\sigma/\lg^{3/2} n + \sqrt{n\sigma\lg\sigma}) 
= O(n\lg\sigma/\lg n + \sigma\lg n)$.
% if sqrt(n s ls) > n ls / ln, then s > n ls / ln^2, or n ls < s ln^2
% then sqrt(n s ls) < sqrt(s^2 ln^2) = s ln

When we store the miniblocks in compressed form, in 
Section~\ref{sec:rrr}, they could use as little as 
$O(\lg^\eps n\lg\lg n)$ bits, and thus we could store up to 
$\Theta(\lg^{1-\eps} n/\lg\lg n)$ miniblocks in a single leaf of $T$. 
This can still can be handled in constant time using (more complicated)
universal tables \cite{MN08}, and the counters and pointers of $O(\lg\lg n)$ 
bits are still large enough.

\subsection{Structure $R_j(v)$}

To support $\ra$ and $\sel$ we enrich $T$ with further information
per node. We store $\rho$ counters with the number
of occurrences of each symbol in the subtree of each child. The node size 
becomes $O(\tau\rho\lg\lg n) = O(\lg^{\eps+\delta}n\lg\lg n) = o(\lg n)$
as long as $\eps+\delta < 1$. This adds up to 
$O(|G_j(v)|\rho \lg\lg n / \lg n)$ bits because the leaves of $T$
handle $\tau$ miniblocks.

With this information on the nodes we can easily solve $\ra$ and $\sel$ in
constant time, by descending on $T$ and determining the correct child
(and accumulating data on the leftward children) in $O(1)$ time using
universal tables. Nodes can also be updated in constant time even upon splits
and merges, since all the counters can be recomputed in $O(1)$ time.

\subsection{Structure $F_j(v)$}

This structure stores all the inter-node pointers leaving from 
block $G_j(v)$, to its parent and to any of the $\rho$ children of node $v$.

The structure is a tree $T_f$ very similar in spirit to $T$. The pointers
are stored at the leaves of $T_f$, in increasing order of their source
position inside $G_j(v)$.
The pointers stored are inter-node, and
thus require $\Theta(\lg n)$ bits. Thus we store a constant number of pointers 
per leaf of $T_f$. For each pointer we store the position in $G_j(v)$ holding the pointer
(relative to the starting position of the leaf node inside $G_j(v)$) 
and the target position (as an absolute pointer to another $G_\ell(u)$). 
The internal nodes, of arity $\tau$, maintain 
information on the number of positions of $G_j(v)$ covered by each child,
and the number of pointers of each kind ($1+\rho$ counters) stored in the 
subtree of each child. This requires $O(\tau\rho\lg\lg n) = o(\lg n)$ bits, as
before. To find the last position before $i$ holding a pointer of a certain
kind (parent or $t$-th wavelet tree child, for any $1\le t\le\rho$), we 
traverse $T_f$ from the root looking for position $i$. At each node $x$, it 
might be that the child $y$ where we have to enter holds pointers of that kind,
or not. If it does, then we first enter into child $y$. If we return with an 
answer, we recursively return it. If we return with no answer, or there are
no pointers of the desired kind below $y$, we enter into the last sibling to 
the left of $y$ that holds a pointer of the desired kind, and switch to a 
different mode where we simply go down the tree looking for the rightmost 
child with a pointer of the desired kind. It is not hard to see that this 
procedure visits $O(1/\delta)$ nodes, and thus it is constant-time because all 
the computations inside nodes can be done in $O(1)$ time with universal tables.
When we arrive at the leaf, 
we scan for the desired pointer in constant time.

The tree $T_f$ must be updated when a symbol $t$ is inserted before any other 
occurrence of $t$ in $G_j(v)$, when a symbol is inserted at the first position 
of $G_j(v)$ and, similarly, when symbols are deleted from $G_j(v)$. The needed
queries are easily answered with tree $T$. Moreover, due to the 
bidirectionality, we must also update $T_f$ when pointers to $G_j(v)$ are 
created from the parent or a child of $v$, or when they are deleted. 
Those updates work just like on the tree
$T$. $T_f$ is also updated upon insertions and deletions of symbols, even 
if they do not change pointers, to maintain the positions up to date. In this
case we traverse $T_f$ looking for the position of the update, change the 
offsets stored at the leaf, and update the subtree sizes stored at the nodes. 

\subsection{Structure $H_j(v)$} 

This structure manages the inter-node pointers that point inside $G_j(v)$.
As explained in Section~\ref{sec:H}, we give a handle to the outside nodes, 
that does
not change over time, and $H_j(v)$ translates handles to positions in $G_j(v)$.

We store a tree $T_h$ that is just like $T_f$, where the incoming pointers 
are stored. $T_h$ is simpler, however, because at each node we only need to
store the number of positions covered by the subtree of each child. It must
also be possible to traverse $T_h$ from a leaf to the root. 

In addition, we manage a table $Tbl$ so that $Tbl[h]$ points to the leaf of 
$T_h$ where the pointer corresponding to handle $h$ is stored. $Tbl$ is also
managed as a tree similar to $T_f$, with pointers sorted by id,
where a constant number of ids $h$ are stored
at the leaves together with their pointers to the leaves of $T_h$ (note that 
there are $O(\lg^3 n /\lg\rho)$ ids at most, so we need $O(\lg\lg n)$ 
bits for both ids and their pointers to $T_h$). Each internal node in $Tbl$ 
maintains the maximum id stored at its leaves and the number of ids stored at 
its leaves. Thus one can in constant time find the pointer to $T_h$ 
corresponding to a given id, and also find the smallest unused id when a
fresh one is needed (by looking for the first leaf of $Tbl$ where the maximum
id is larger than the number of ids).

At the leaves of $T_h$ we store, for each pointer, a backpointer 
to the corresponding leaf of $Tbl$ and the position in $G_j(v)$ (in relative 
form). Given a handle $h$, we find using $Tbl$ the corresponding position in the
leaf of $T_h$, and move upwards up to the root of $T_h$, adding to the leaf 
offset the 
number of positions covered by the leftward children of each node. At the end 
we have obtained the position in constant time.

When pointers to $G_j(v)$ are created or destroyed, we insert or remove
pointers in $T_h$. This requires traversing it top-down to find the appropriate
leaf position and returning back to the root updating offsets. Backpointers to
$Tbl$ are used to adjust a constant number of positions in the leaf of $T_h$.
We must also update $T_h$ upon symbol insertions and deletions in
$G_j(v)$, to maintain the positions up to date. When a leaf splits or merges,
we also update the pointers from a constant number of positions in
$Tbl$, found with the backpointers. Similarly, the insertion and deletion of
pointers from outside require updating $Tbl$, and the backpointers from $T_h$
are maintained up to date using the pointers from $Tbl$ to $T_h$.

$Tbl$ may contain up to $\Theta(\lg^3 n/\lg\rho)$ pointers of $O(\lg\lg n)$ 
bits, which can be significant for some blocks. However, across the whole
structure there can be only $O(\rho n\lg\sigma/\lg^3 n)$ pointers,
adding up to $s=O(\rho n\lg\sigma \lg\lg n/\lg^3 n)$ bits, spread across
$a=O(n\lg\sigma/\lg^3 n)$ tables $Tbl$. Using again Lemma~\ref{lem:EA},
a collection of EAs poses an overhead of $O(n\lg\sigma/\lg^2 n)$ bits.

\subsection{Structure $D_j(v)$ and the Final Result}

Structure $D_j(v)$ is implemented as a tree $T_d$ analogous to $T$, storing at 
each node the number 
of positions and the number of non-deleted positions below each child. It
requires $O(|G_j(v)|\lg\lg n / \lg n)$ bits. Since these are stored only for
the root $v_r$ and the leaves $v_a$ of $\mathcal{T}$, its space adds up to
$O(n\lg\rho \lg\lg n / \lg n) = O(n (\lg\lg n)^2/\lg n)$ bits.

While the raw data adds up to $n\lg\sigma$ bits, the space overhead adds up
to $O(n\lg\sigma \lg^\eps n\lg\lg n/\lg n)$ for all the pointers plus
$O(n\lg\sigma/\lg n+\sigma \lg n)$ for the memory management overhead. 
We can use, say, $\delta = \eps$ and then have $O(1/\eps)$ time and
$O(n\lg\sigma/\lg^{1-\eps} n+\sigma\lg n)$ bits for any 
$0<\eps<1$ (renaming $2\eps$ as $\eps$).
However, when the data is compressed (Section~\ref{sec:rrr}), the sum of all the
$|G_j(v)|$ terms in the space is $nH_0(S)+O(n\lg\sigma\lg\lg n/\lg^\eps n)$.
This makes the space overhead related to the memory management and of
$R_j(v)$ structures add up to $O(nH_0(S)/\lg^{1-\eps}n + n\lg\sigma/\lg n +
\sigma\lg n)$
bits.

\section{Extensions and Applications}
\label{sec:app}

We first describe an extension of our results to handling general alphabets,
and then various applications of the original and the extended results.

\subsection{Handling General Alphabets}
\label{sec:alphabet}

Our time results do not depend on the alphabet size $\sigma$, yet our space 
does, in a way that ensures that $\sigma$ gives no problems as long as 
$\sigma=o(n)$ (so $\sigma\lg n = o(n\lg\sigma)$).

Let us now consider the case where the alphabet $\Sigma$ is much larger than
the {\em effective} alphabet of the string, that is, the set of symbols that 
actually appear in $S$ at a given point in time. Let us now use $s\le n$
to denote the effective alphabet size. Our aim is to maintain the space within 
$nH_0(S) + o(n\lg s) + O(s\lg n)$ 
bits, even when the symbols come from a large universe $\Sigma=[1..|\Sigma|]$, 
or even from a general ordered universe such as $\Sigma = \mathbb{R}$ or
$\Sigma = \Gamma^*$ (i.e., $\Sigma$ are words over another
alphabet $\Gamma$).

Our mappings $SN$ and $NS$ of Section~\ref{sec:space} 
give a simple way to handle a 
sequence over an unbounded ordered alphabet. By changing $SN$
to a custom structure to search $\Sigma$, and storing elements of 
$\Sigma$ in array $NS$, we obtain the following results, using respectively
Theorems~\ref{thm:compr} and \ref{thm:worstcase}.

\begin{theorem}
\label{thm:general}
A dynamic string $S[1,n]$ over a general alphabet $\Sigma$ 
can be stored in a structure using 
$nH_0(S) + o(n\lg s) + O(s\lg n) + \mathcal{S}(s)$ 
bits and supporting queries
$\acc$, $\ra$ and $\sel$ in time 
$O(\mathcal{T}(s)+\lg n/\lg \lg n)$.
Insertions and deletions of symbols 
are supported in $O(\mathcal{U}(s)+\lg n/\lg\lg n)$ amortized 
time. 
Here $s\le n$ is the number of distinct symbols of $\Sigma$ occurring in $S$,
$\mathcal{S}(s)$ is the number of bits used by a dynamic data structure 
to search over $s$ elements in $\Sigma$ plus to refer to $s$ 
elements in $\Sigma$, $\mathcal{T}(s)$ is the worst-case time to search 
for an element among $s$ of them in $\Sigma$, and $\mathcal{U}(s)$
is the amortized time to insert/delete symbols of $\Sigma$ in the structure.
\end{theorem}

\begin{theorem}
\label{thm:generalwc}
A dynamic string $S[1,n]$ over a general alphabet $\Sigma$ 
can be stored in a structure using 
$nH_0(S) + o(n\lg s) + O(s\lg n) + \mathcal{S}(s)$ 
bits and supporting queries
$\acc$ and $\sel$ in time $O(\mathcal{T}(s)+\lg n/\lg \lg n)$ and
$\ra$ in time $O(\mathcal{T}(s)+\lg n)$.
Insertions and deletions of symbols 
are supported in $O(\mathcal{U}(s)+\lg n)$ time. 
Here $s\le n$ is the number of distinct symbols of $\Sigma$ occurring in $S$,
$\mathcal{S}(s)$ is the number of bits used by a dynamic data structure 
to search over $s$ elements in $\Sigma$ plus to refer to $s$ 
elements in $\Sigma$, $\mathcal{T}(s)$ is the time to search 
for an element among $s$ of them in $\Sigma$, and $\mathcal{U}(s)$
is the time to insert/delete symbols of $\Sigma$ in the structure.
All times are worst-case
\end{theorem}

Using general and dynamic alphabets had not been achieved in previous 
dynamic sequence data structures, because the wavelet has a static shape
(and changing it is costly). These results open the door to using
these solutions in various scenarios where alphabet dynamism is essential.
We examine a few interesting particular cases:

\begin{itemize} 
\item 
We can handle a sequence of arbitrary real numbers in the comparison model,
by using a balanced tree for the alphabet data structure.
If $\Sigma=\mathbb{R}$ we have $O(\lg s+\lg n / \lg\lg n)$
times using Theorem~\ref{thm:general} and $O(\lg n)$ worst-case times using
Theorem~\ref{thm:generalwc}. Those complexities are optimal in the comparison
model.
\item 
We can handle a sequence of strings, that is,
$\Sigma=\Gamma^*$ on a general alphabet
$\Gamma$. Here we can store the effective set of strings in a data structure 
by Franceschini and Grossi \cite{FG04}, so that operations involving a string 
$a$ take $O(|a|+\lg\gamma + \lg n / \lg\lg n)$,
where $\gamma$ is the number of symbols of $\Gamma$ actually in use.
With Theorem~\ref{thm:generalwc} we obtain worst-case times 
$O(|a|+\lg\gamma+\lg n)$.
\item
If $\Sigma=[1..|\Sigma|]$ is a large integer range, we can obtain time
$O(\lg\lg|\Sigma| + \lg n / \lg\lg n)$, or worst-case times 
$O(\lg\lg|\Sigma|+\lg n)$, and the space increases by $O(s\lg|\Sigma|)$ bits, 
by using y-fast tries \cite{Wil83} to handle the alphabet.
\item
Another important particular case is when we maintain a contiguous effective
alphabet $[1..s]$, and only insert new symbols $\sigma+1$. This is the case
where the symbol identities themselves are not important. In this case
there is no time penalty for letting the alphabet grow dynamically.
\end{itemize}

\no{
\subsection{Block Updates}

\section{Applications} \label{sec:app}

Our new results impact in a number of applications that build on dynamic
sequences. We describe several here.
}

\subsection{Dynamic Sequence Collections} 

A landmark application of dynamic
sequences, stressed out in several papers along time
\cite{CHL04,MN06,CHLS07,MN06,LP07,MN08,GN08,LP09,GN09,HM10,NS10}, 
is to maintain a collection $\mathcal{C}$ of
texts, where one can carry out indexed pattern matching, as well as inserting
and deleting texts from the collection. Plugging in our new representation we 
can significantly improve the time and space of previous work, with an amortized and with
a worst-case update time, respectively.

\begin{theorem} \label{thm:fmindex}
There exists a data structure for handling a collection $\mathcal{C}$ of
texts over an alphabet $[1..\sigma]$ within size
$nH_h(\mathcal{C})+o(n\lg\sigma)+
 O(\sigma^{h+1}\lg n + m\lg n)$ bits, 
simultaneously for all $h$. 
Here $n$ is the length of the concatenation
of $m$ texts, 
 $\mathcal{C}=\ T_1 \circ \ T_2 \cdots$ $\circ\ T_m$, and we assume that
the alphabet size is $\sigma=o(n)$.
The structure supports counting of the occurrences
of a pattern $P$ in $O(|P|\lg n/\lg\lg n)$ time.
After counting, any occurrence can be located in time $O(\lg_\sigma n\lg n)$. 
Any substring of length $\ell$ from any $T$ in the collection can be displayed 
in time $O((\ell/\lg\lg n + \lg_\sigma n) \lg n)$.
Inserting or 
deleting a text $T$ takes $O(\lg n + |T|\lg n/\lg\lg n)$ amortized time.
For $0 \le h \le (\alpha \lg_\sigma n)-1$, for any constant $0<\alpha<1$, the 
space simplifies to $nH_h(\mathcal{C})+o(n\lg\sigma)+O(m\lg n)$ bits.
\end{theorem}

\begin{theorem} \label{thm:fmindexwc}
There exists a data structure for handling a collection $\mathcal{C}$ of
texts over an alphabet $[1..\sigma]$ within size
$nH_h(\mathcal{C})+o(n\lg\sigma)+
 O(\sigma^{h+1}\lg n + m\lg n)$ bits, 
simultaneously for all $h$. 
Here $n$ is the length of the concatenation
of $m$ texts, 
 $\mathcal{C}=\ T_1 \circ \ T_2 \cdots$ $\circ\ T_m$, and we assume that
the alphabet size is $\sigma=o(n)$.
The structure supports counting of the occurrences
of a pattern $P$ in $O(|P|\lg n)$ time.
After counting, any occurrence can be located in time $O(\lg_\sigma n\lg n\lg\lg n)$. 
Any substring of length $\ell$ from any $T$ in the collection can be displayed 
in time $O((\ell + \lg_\sigma n \lg\lg n) \lg n)$.
Inserting or 
deleting a text $T$ takes $O(|T|\lg n)$ time.
For $0 \le h \le (\alpha \lg_\sigma n)-1$, for any constant $0<\alpha<1$, the 
space simplifies to $nH_h(\mathcal{C})+o(n\lg\sigma)+O(m\lg n)$ bits.
\end{theorem}

The theorems refer to $H_h(\mathcal{C})$, the $h$-th order empirical 
entropy of sequence $\mathcal{C}$ \cite{Man01}. This is a lower bound to any
semistatic statistical compressor that encodes each symbol as a function of
the $h$ preceding symbols in the sequence, and it holds $H_h(\mathcal{C}) \le
H_{h-1}(\mathcal{C}) \le H_0(\mathcal{C}) \le \lg\sigma$ for any $h>0$.
To offer search capabilities, the Burrows-Wheeler Transform (BWT)
\cite{BW94} of $\mathcal{C}$, $\mathcal{C}^{bwt}$, is represented, not 
$\mathcal{C}$; then $\acc$ and $\ra$ operations on $\mathcal{C}^{bwt}$ 
are used to support pattern
searches and text extractions. K\"arkk\"ainen and Puglisi \cite{KP11} showed 
that, if 
$\mathcal{C}^{bwt}$ is split into superblocks of size $\Theta(\sigma\lg^2 n)$,
and a zero-order compressed representation is used for each superblock, the
total bits are $nH_h(\mathcal{C}) +o(n)$.

We use their partitioning, and Theorems~\ref{thm:compr} or
\ref{thm:worstcase} to represent each superblock. 
For Theorem~\ref{thm:fmindex}, the superblock sizes are easily maintained upon 
insertions and deletions of symbols, by splitting and merging superblocks and 
rebuilding the structures involved, without affecting the amortized time per 
operation. They \cite{KP11} also need to manage a table storing the rank of 
each symbol up to the beginning of each superblock. This is arranged, in the 
dynamic scenario, with $\sigma$ partial sum data structures containing 
$O(n/(\sigma\lg^2 n))$ elements each, plus another one storing the superblock 
lengths. This adds $O(n/\lg n)$ bits and $O(\lg n / \lg\lg n)$ time per 
operation \cite[Lem.~1]{NS10}. Upon blocks splits and merges, we use the
same techniques used for $\cP$ structures described in Section~\ref{sec:H}.

For Theorem~\ref{thm:fmindexwc} we use the smooth block size management
algorithm described in Section~\ref{sec:indels} for the superblocks, which 
guarantees worst-case times and the same space redundancy. Then partial-sum 
data structures are used without problems.

Finally, the locating and displaying overheads are obtained by marking one
element out of $\lg_\sigma n\lg\lg n$, so that the space overhead of 
$o(n\lg\sigma)$ is maintained. Other simpler data structures used in previous
work \cite{MN08}, such as mappings from document identifiers to their position
in $\mathcal{C}^{bwt}$ and the samplings of the suffix array, can easily be 
replaced by $O(\lg n / \lg\lg n)$ time partial-sums data structures and simpler
structures to maintain dictionaries of values \cite[Lem.~1]{NS10}.

\subsection{Burrows-Wheeler Transform} 

Another application
of dynamic sequences is to build the BWT of a text $T$, $T^{bwt}$, within
compressed space, by starting from an empty sequence and inserting each new
character, $T[n]$, $T[n-1]$, $\ldots$, $T[1]$, at the proper positions. 
Equivalently, this corresponds to initializing an empty collection and then
inserting a single text $T$ using Theorem~\ref{thm:fmindex}. The
result is also stated as the compressed construction of a static FM-index
\cite{FMMN07}, a compressed index that consists essentially of a (static) 
wavelet tree of $T^{bwt}$. Our new representation improves upon the best 
previous result on compressed space \cite{NS10}.

\begin{theorem}
The Alphabet-Friendly FM-index \cite{FMMN07}, as well as the BWT \cite{BW94}, 
of a text $T[1,n]$ over an alphabet of size $\sigma$, can be built using 
$nH_h(T)+o(n\lg\sigma)$ bits, simultaneously for
all $1 \le h \le (\alpha \lg_\sigma n)-1$ and any constant $0<\alpha<1$, in time
$O(n\lg n/\lg\lg n)$. It can also be built within the same time and 
$nH_0(T)+o(n\lg\sigma)+O(\sigma\lg n)$ bits, for
any alphabet size $\sigma$.
\end{theorem}

We are using Theorem~\ref{thm:fmindex} for the case $h>0$, and 
Theorem~\ref{thm:compr} to obtain a less alphabet-restrictive result for
$h=0$ (in this case, we do not split the text into superblocks of 
$O(\sigma\lg^2 n)$ symbols, but just use a single sequence). Note that,
although insertion times are amortized in those theorems, this result is
worst-case because we compute the sum of all the insertion times.

This is the first time that $o(n\lg n)$ time
complexity is obtained within compressed space. Other
space-conscious results that achieve better time complexity (but more space)
are Okanohara and Sadakane \cite{OS09}, who achieved optimal 
$O(n)$ time within $O(n \lg \sigma \lg\lg_\sigma n)$ bits, and
Hon et al.\ \cite{HSS09}, who achieved $O(n\lg\lg\sigma)$ time and 
$O(n\lg\sigma)$ bits. Older results, like K\"arkk\"ainen's \cite{Kar07}, 
are superseded.

\subsection{Binary Relations}

Barbay et al.\ \cite{BGMR07} show how to represent a binary relation of $t$ pairs
relating $n$ ``objects'' with $\sigma$ ``labels'' by means of a string of
$t$ symbols over alphabet $[1..\sigma]$ plus a bitmap of length $t+n$. The
idea is to traverse the matrix, say, object-wise,
and write down in a string the labels of the pairs found. Meanwhile we append
a 1 to the bitmap each time we find a pair and a 0 each time we move to the
next object. Then queries like: find the objects related to a label, find the
labels related to an object, and tell whether an object and a label are
related, are answered via access, rank and select operations on the string
and the bitmap.

A limitation in the past to make this representation dynamic was that creating
or removing labels implied changing the alphabet of the string. Now we can use
Theorem~\ref{thm:general} to obtain a fully dynamic representation. We 
illustrate the case where labels and objects are contiguous values in integer 
intervals $[1..\sigma]$ and $[1..n]$, respectively. We note that the structure
on the sequence of labels is so fast that the bitmap, which is longer, dominates
the times.

\begin{theorem} \label{thm:binrel}
A dynamic binary relation consisting of $t$ pairs relating $n$ objects 
with $\sigma$ labels can support the operations of 
counting and listing the objects 
related to a given label, counting and listing the labels related to a given 
object, and telling whether an object and a label are related, all in time 
$O(\lg (n+t)/\lg\lg (n+t))$ per delivered datum. Pairs, objects 
and labels can also be added and deleted in amortized time 
$O(\lg (n+t) / \lg\lg (n+t))$. The space required is 
$tH + o(t\lg\sigma)+n\lg n+\sigma\lg \sigma+ O(t+n+\sigma\lg t)$ bits,
where $H=\sum_{1 \le i \le \sigma} (t_i/t)\lg(t/t_i)
\le \lg\sigma$, where $t_i$ is the number of objects related to label $i$.
Only labels and objects with no related pairs can be deleted.
\end{theorem}

\begin{theorem} \label{thm:binrelwc}
A dynamic binary relation consisting of $t$ pairs relating $n$ objects 
with $\sigma$ labels can support the operations of 
counting and listing the objects 
related to a given label, counting and listing the labels related to a given 
object, and telling whether an object and a label are related, all in time 
$O(\lg (n+t))$ per delivered datum. Pairs, objects 
and labels can also be added and deleted in time 
$O(\lg (n+t))$. The space required is 
$tH + o(t\lg\sigma)+n\lg n+\sigma\lg \sigma+ O(t+n+\sigma\lg t)$ bits,
where $H=\sum_{1 \le i \le \sigma} (t_i/t)\lg(t/t_i)
\le \lg\sigma$, where $t_i$ is the number of objects related to label $i$.
Only labels and objects with no related pairs can be deleted.
\end{theorem}

The careful reader may notice that we have uniformized the times of all the
operations for simplicity, yet some can be slightly faster. For example,
listing the $m$ labels related to a given object requires only
$O(\lg(n+t)/\lg\lg(n+t)+m\lg t/\lg\lg t)$ time. Also, obviously, we can
exchange labels and objects if desired.

\subsection{Directed Graphs} 

A particularly interesting and general binary
relation is a directed graph with $n$ nodes and $e$ edges. Our binary relation
representation allows one to navigate a directed graph in forward and backward 
direction, and modify it, within about the space needed by a classical
adjacency list representation, and even less.

\begin{theorem}
A dynamic directed graph consisting of $n$ nodes and $e$ edges 
can support the operations of counting and listing the neighbors pointed from 
a node, counting and listing the reverse neighbors pointing to a node, and 
telling whether there is a link from one node to another, all in time 
$O(\lg (n+e) / \lg\lg (n+e))$ per delivered datum. Nodes and edges 
can be added and deleted in amortized time 
$O(\lg (n+e) / \lg\lg (n+e))$. The space used is 
$eH + o(e\lg n) + n\lg n + O(e+n\lg e)$ bits, 
where $H=\sum_{1 \le i \le n} (e_i/e)\lg(e/e_i) \le \lg n$
and $e_i$ is the outdegree of node $i$.
\end{theorem}

\begin{theorem}
A dynamic directed graph consisting of $n$ nodes and $e$ edges 
can support the operations of counting and listing the neighbors pointed from 
a node, counting and listing the reverse neighbors pointing to a node, and 
telling whether there is a link from one node to another, all in time 
$O(\lg (n+e))$ per delivered datum. Nodes and edges 
can be added and deleted in time 
$O(\lg (n+e))$. The space used is 
$eH + o(e\lg n) + n\lg n + O(e+n\lg e)$ bits, 
where $H=\sum_{1 \le i \le n} (e_i/e)\lg(e/e_i) \le \lg n$
and $e_i$ is the outdegree of node $i$.
\end{theorem}

Note also that we can change ``outdegree'' by ``indegree'' in the theorem by
representing the transposed graph, as operations are symmetric. Our ability
to handle dynamic alphabets is essential here to allow node insertions and
deletions in the graph.

\subsection{Inverted Indexes}

Finally, we consider an application where the symbols are strings. Take a
text $T$ as a sequence of $n$ {\em words}, which are strings over a set of
letters $\Gamma$. The alphabet $\Gamma$ is integer and fixed, of size $\gamma$. 
The alphabet of $T$ is $\Sigma = \Gamma^*$, and its effective
alphabet is called the {\em vocabulary} $V$ of $T$, of size $|V|=\sigma$.
A {\em positional inverted index} is a data structure that, given a word
$w \in V$, returns the positions in $T$ where $w$ appears \cite{BYRN11}.

A well-known way to simulate a positional inverted index within no extra space
on top of the compressed text is to use a compressed sequence representation 
for $T$ (over alphabet $\Sigma$), so that operation $\sel_w(T,i)$ simulates
access to the $i$th position of the list of word $w$, whereas access to the
original $T$ is provided via $\acc(T,i)$. Operation rank can be used to 
emulate various inverted index algorithms, particularly for intersections
\cite{BN09}. The space is the zero-order entropy of the text seen as a
sequence of words, which is very competitive in practice \cite{BYRN11}. 
Our new technique
permits modifying the underlying text, that is, it simulates a dynamic
inverted index. For this sake we use Theorem~\ref{thm:general}
and compact tries to handle a vocabulary over a fixed alphabet.

\begin{theorem}
A text of $n$ words with a vocabulary of $\sigma$ words and total length $\nu$
over a fixed alphabet $\Gamma$ of size $\gamma$ can be represented 
within $nH_0(T) + o(n\lg\sigma) + O(\nu\lg\gamma+\sigma\lg n)$
bits of space, where $H_0(T)$ is
the word-wise entropy of $T$.
The representation outputs any word $T[i]=w$
given $i$, finds the position of the $i$th occurrence of any word $w$,
and tells the number of occurrences of any word $w$ up to position $i$, all
in time $O(|w|+\lg n / \lg\lg n)$. A word $w$ can be inserted or deleted at 
any position in $T$ in amortized time $O(|w|+\lg n/\lg\lg n)$.
\end{theorem}

\begin{theorem}
A text of $n$ words with a vocabulary of $\sigma$ words and total length $\nu$
over a fixed alphabet $\Gamma$ of size $\gamma$ can be represented 
within $nH_0(T) + o(n\lg\sigma) + O(\nu\lg\gamma+\sigma\lg n)$
bits of space, where $H_0(T)$ is
the word-wise entropy of $T$.
The representation outputs any word $T[i]=w$
given $i$, and finds the position of the $i$th occurrence of any word $w$,
in time $O(|w|+\lg n / \lg\lg n)$.
It tells the number of occurrences of any word $w$ up to position $i$, and
supports the insertion or deletion of any word $w$ in $T$, in time 
$O(|w|+\lg n)$.
\end{theorem}

We remark that $\sigma$ and $\nu$ are assumed to be $O(n^\alpha)$ for some 
$0<\alpha<1$ in information retrieval models \cite{BYRN11}. Under this
assumption the space is just $nH_0(T) + o(n\lg\sigma)$.

Another kind of inverted index, a {\em non-positional} one, relates each word
with the documents where it appears (not to the exact positions). This can be
seen as a direct application of our binary relation representation
\cite{BCN10}, and our dynamization theorems apply to it as well.

\section{Conclusions and Further Challenges} \label{sec:concl}

We have obtained $O(\lg n /\lg\lg n)$ time for all the operations that handle 
a dynamic sequence on an arbitrary alphabet $[1..\sigma]$, matching lower 
bounds that apply to binary alphabets \cite{FS89}, and using zero-order
compressed space. Our structure is faster than the best previous work 
\cite{HM10,NS10} by a factor of $\Theta(\lg\sigma/\lg\lg n)$ when the
alphabet is larger than polylogarithmic. 
The query times are worst-case, yet the update times are amortized. We also
show that it is possible to obtain worst-case for all the operations, although
times for $\ra$ and updates raises to $O(\lg n)$.
We also show how to handle general and infinite alphabets.
Our result can be applied to a number of problems and improve previous
upper bounds on those; we have described several ones.

We remark that the lower bounds \cite{FS89} are valid also for amortized times,
so our amortized solution is optimal, yet it is not known whether our
worst-case solution is optimal. Thus the main remaining challenge is whether 
it is possible to attain the optimal $O(\lg n / \lg\lg n)$ worst-case time 
for all the operations. 

Another interesting challenge is to support a stronger set of update 
operations, such as block edits, concatenations and splits in the sequences.
Navarro and Sadakane \cite{NS10} support those operations within time
$O(\sigma\lg^{1+\eps} n)$. While it seems feasible to achieve, in our 
structure, $O(\sigma\lg n)$ time by using blocks of $\Theta(\lg^2 n)$ bits,
the main hurdle is the difficulty of mimicking the same splits and
concatenations on the list maintenance data structures we use 
\cite{IA84,Mor03}.

%OJO can we obtain $O(\lg n / \lg w)$ time by doing bit magic in the blocks?

\bibliographystyle{alpha}
\bibliography{paper}

\end{document}